\documentclass[11pt,a4paper]{article}

\usepackage{polski}
\usepackage[english]{babel}

\usepackage{jheppub}
\usepackage{amsthm,amssymb,amsmath,epic,float}
\usepackage{rotating,indentfirst,array,varioref}
\usepackage{appendix,tikz,mathtools}
\usepackage{setspace}
\usepackage{longtable}
\usepackage{tikz}
\usetikzlibrary{decorations.pathreplacing}
\usetikzlibrary{calc}
\usepackage{enumitem}
\usepackage{hyphenat}
\usetikzlibrary{positioning}
\usepackage{graphicx}  
\usepackage{adjustbox}

\usepackage{arydshln}
\usepackage{cancel}

\usepackage{subfig}
\usepackage{comment}
\usetikzlibrary{intersections}
\pgfdeclarelayer{background}
\pgfsetlayers{background,main}

\usepackage{tikz,pgf}
\usetikzlibrary{shapes}
\usetikzlibrary{calc}
\usetikzlibrary{decorations.pathmorphing}
\usetikzlibrary{decorations.pathreplacing,shapes.misc}
\usetikzlibrary{positioning}
\usetikzlibrary{arrows}
\usetikzlibrary{decorations.markings}
\usetikzlibrary{shadings}
\usetikzlibrary{intersections}

\def\be{\begin{equation}}
\def\ee{\end{equation}}

\def\be{\begin{equation}}
\def\en{\end{equation}}
\def\ber{\begin{eqnarray}}
\def\enr{\end{eqnarray}}

\newcommand{\ket}[1]{\left| #1 \right\rangle}

\newcommand{\pd}{\partial}

\newcommand{\br}[1]{{\overline{#1}}}

\def\<{\left(}
\def\>{\right)}

\usepackage{jheppub}
\makeatletter
\def\@fpheader{\vspace{-.1cm}}
\makeatother

\title{\boldmath $(2,2p+1)$ minimal string and intersection theory I}

\author[a,b]{A.~Artemev,}
\author[a,c]{I.~Chaban}

\affiliation[a]{Krichever Center, Skolkovo Institute of Science and Technology, 121205, Moscow, Russia}
\affiliation[b]{Landau Institute for Theoretical Physics, 142432, Chernogolovka, Russia}
\affiliation[c]{National Research University Higher School of Economics, Moscow, Russia}

\emailAdd{artemev.aa@phystech.edu}

\abstract{In view of the recent progress in studying matrix model-2D gravity duality, we reexamine some features of $(2,2p+1)$ minimal string. After reviewing both sides of the proposed correspondence in this case, a previously unnoted identification between correlation numbers of tachyon operators in certain domain of parameter space and "$p$-deformed volumes", which are certain integral transforms of topological recursion data, is described and clarified. This identification allows us to efficiently study correlation numbers at finite matter central charge. In particular, we obtain an intersection-theoretic formula and the simplest recurrent equations for them, analogous to the ones recently derived for Virasoro minimal string. These formulas might be useful in establishing a more thorough connection between worldsheet and matrix model approaches.}


\begin{document} 
\maketitle
\flushbottom
\section{Introduction}
Minimal string (MS) is one of exactly solvable models of non-critical string theory, or two-dimensional gravity, extensively studied since the 1980s. An interesting thing about it is a conjectured duality to a double-scaling limit of certain matrix models (see e.g. \cite{franc1995} for the review). So far, there is a lot of evidence for this conjecture, although it has not been proven in full since it is difficult to perform analytic calculations on the worldsheet. The dual description is convenient because calculation of perturbative string correlators is significantly more streamlined (up to some subtleties concerning the precise matching of correlators \cite{Moore:1991ir}, \cite{belzam2009}). It also allows to address questions about non-perturbative physics of the theory (see e.g. \cite{Gregori:2021tvs}, \cite{Eniceicu:2022dru}).

In some sense the minimal string can be regarded as a deformation of Witten-Kontsevich topological gravity \cite{Witten:1990hr}, for which there are at least three formulations: using field theory (with a certain $c=-2$ worldsheet matter CFT), matrix models or intersection theory of tautological classes on the moduli space of stable curves. One can not, however, immediately generalize the methods used in establishing these equivalent formulations to different non-critical string theories, including minimal string, for which a matrix model description is lacking direct worldsheet derivation. Despite that, more recent studies (in particular, studies of the semiclassical limit of large matter central charge) reveal that in minimal string correlators also have some interpretation in terms of moduli space geometry (\cite{Artemev_2022}, \cite{eberhardt20232d}).

Another nontrivial theory was discovered recently; it is the so-called Virasoro minimal string (VMS, \cite{Collier:2023cyw}). The worldsheet CFT for this theory consists of Liouville CFT of central charge $c>25$; ``timelike'', or ``$c<1$  Liouville  CFT'' \cite{clessthan12015} as ``matter'' and BRST ghosts. Exploring previously proposed links with quantization of Teichmuller space, the authors of \cite{Collier:2023cyw} managed to find an intersection-theoretic formula for correlators in this theory which serves as a bridge between worldsheet and double-scaled matrix model descriptions. Attempt on generalizing this work to better understand the correspondence between worldsheet CFT and matrix models for minimal string is the main motivation for our study.

An algorithm producing $(2, 2p+1)$ MS correlation numbers was proposed in \cite{belzam2009} via matrix model approach (see also \cite{beltar2010}). The proposal passes all available tests from the worldsheet formulation; namely, it satisfies CFT fusion rules and agrees with direct calculations \cite{Belavin:2005jy}, \cite{Artemev:2022sfi}. The semiclassical limit $p\to \infty$ of these correlators was investigated in \cite{Artemev_2022}, \cite{eberhardt20232d}. This allows to elucidate the geometric meaning of correlation numbers in this limit as “moduli space volumes for cone surfaces”.

The complicated and quite bulky structure of expressions in \cite{belzam2009} might cause difficulty for further analysis.  One of the key results of our paper is the relation (\ref{cornumandpdef})
\begin{equation*}
\left. \frac{\pd^n \mathcal{F}_g}{\pd \tau_{k_1} \dots \pd \tau_{k_n}} \right|_{\tau_0 = - \frac{1}{2}, \tau_1 \dots \tau_{p-1} = 0} \sim V_{g,n} \left( \lambda_j = \frac{i}{2}(2(p-k_j)-1) \right) 
\end{equation*}
between correlation numbers proposed in \cite{belzam2009} and previously studied “$p$-deformed volumes” \cite{mertens2021}. The benefit is that the latter have a convenient description in terms of a well-studied framework of topological recursion. In particular this allows us to obtain an expression (\ref{VfromRR})
\begin{equation*}
V_{g, n} \sim
(8 \pi^2 b^2)^{3g - 3 + n}  \int \limits_{\mathcal{\overline{M}}_{g, n}} e^{\frac{c - 13}{24}  \kappa_1 -  \sum \limits_{m \geq 1} \frac{B_{2m}}{(2m)(2m)!}\kappa_{2m} + \sum \limits_{k = 1}^{n}  P_k^2 \psi_{k} + \sum \limits_{k \geq 0}\tilde{b}_{k} \delta_{k, 0}} 
\end{equation*}
for correlation numbers as intersection numbers on moduli spaces of curves.
The method for deriving this formula is essentially the same as for VMS and consists of applying the machinery of \cite{Eynard:2011sc}. The feature of our case is in taking a local change of coordinate on spectral curve before using the formula from \cite{Eynard:2011sc}. After this change of coordinate, VMS and MS spectral curve data differ only in bidifferentials.

We believe that this formula might elucidate the geometric definition of MS amplitudes, akin to VMS description or the one available for Polyakov measure for usual critical (super)string \cite{Belavin:1986cy}, \cite{Voronov:1988ms}. For now we have not found the definite answer to this question.

This paper is structured as follows. In section \ref{cftprelims}, we introduce notations and give some background information on the objects of our interest --- ``correlation numbers'' in minimal string theory, focusing on $(2,2p+1)$ case. In section \ref{toprecprelim}, we describe the previously proposed matrix model analogues of them. Then we come to the main results of this paper: in section \ref{volandnum}, we relate several definitions of ``correlation numbers'' in the ``topological recursion'' language mentioned before. In sections \ref{varchange} and \ref{inttheo} we describe how one can obtain an intersection-theoretic formula for these correlation numbers. Although a general formula for topological recursion data exists for any double-scaled matrix model, we propose and prove a different one, which is, interestingly, very close to the VMS answer. We discuss some possible interpretations of this formula in section \ref{spec}. In section \ref{recequ}, we study what form the simplest ``string'' and ``dilaton'' recurrent equations in the topological recursion framework obtain for correlation numbers and comment on the possible CFT interpretation of these equations. We finish with some concluding thoughts on further directions to study in section \ref{conc}.

\section{Preliminaries}
\subsection{Minimal string worldsheet theory} \label{cftprelims}
\paragraph{Definition and notations. }
Worldsheet CFT for minimal string theory (also known as minimal Liouville gravity) is a CFT of total central charge zero consisting of a CFT minimal model as ``matter'' (we will discuss only $(2,2p+1)$ series), a Liouville CFT and anticommuting $BC$-ghosts of $c_{gh}=-26$
\begin{align}
    &A_{MLG} = A_L + A_{M_{2,2p+1}} +\underbrace{\frac{1}{\pi}\int d^2x\,\left( C \br{\pd} B + \br{C} \pd \br{B} \right)}_{A_{ghost}};\, \\
    &A_L = \int d^2x\,\sqrt{\hat{g}}\left(\frac{1}{4\pi}\hat{g}^{ab} \pd_a \phi \pd_b \phi + \mu e^{2b\phi} + \frac{Q}{4\pi} \hat{R} \phi \right),\,Q \equiv b + b^{-1}
\end{align}
The condition of zero central charge leads to the relation between $p$ and Liouville parameter $b$: $b = \sqrt{2/(2p+1)}$.

To set up notations, we will describe primary operators in Liouville and matter sectors. In $(2,2p+1)$ minimal model, the spectrum consists of $p$ primary fields from the Kac table $\Phi_{1,k},\,k = 1 \dots 2p$ with the additional identification $\Phi_{1,k} = \Phi_{1,2p+1-k}$. Dimension of $\Phi_{1,k}$ is $\Delta^M_{1,k} = \alpha_{1,k} (\alpha_{1,k} - b^{-1}+b)$, where $\alpha_{1,k} = \frac{b}{2}(k-1)$. Virasoro representation, corresponding to each primary field in minimal model, is doubly degenerate. 

In Liouville sector, primary operators are exponentials $V_a = \exp (2 a \phi)$ with an additional identification $V_a = R(a) V_{Q-a}$. Their (bulk) conformal dimension is $\Delta_a^L = a (Q-a)$\footnote{If we consider the boundary operator represented by the same exponential, its dimension is $\Delta_a^{L,b} = 2a(Q-2a)$}. Dependent on circumstances, it may be convenient to parametrize them instead of $a$ via ``Liouville momenta'' $P: a = \frac{Q}{2} + i P$, or, for special values of $a$, with a pair of integers $(m,n): a_{m,n} = \frac{-(m-1) b^{-1} - (n-1) b}{2}$ (positive $m,n$ correspond to degenerate Virasoro representations).
\paragraph{BRST cohomology. }
 From the requirement of zero total central charge it also follows that the theory has a nilpotent ($Q_{\text{BRST}}^2 = 0$) BRST-symmetry with generator
\begin{equation}
Q_{\text{BRST}} \equiv \mathcal{Q} + \br{\mathcal{Q}};\,\mathcal{Q} \equiv \oint dz\, \left(C (T^L + T^M) + :C \pd C B: \right)
\end{equation}
and an analogous expression for antiholomorphic charge $\mathcal{Q}$. Equivalently one can write a mode expansion
\begin{equation}
\mathcal{Q} = \sum \limits_n c_n (L_{-n} - \delta_{n,0}) - \frac{1}{2} \sum \limits_{n,m} (m-n) :c_{-m} c_{-n} b_{n+m}: \label{brstc}
\end{equation}
String theoretic observables are given by cohomology classes of this BRST operator; these are additionally graded by their ghost number. In minimal string theory the classes represented by (bulk) local operators were fully classified by Lian and Zuckerman in \cite{lianzuck}. It turns out that (in $(2,2p+1)$ case) there is one such cohomology class with a definite ghost number for each Liouville parameter $a = a_{1,-n}; n \in \mathbb{Z}, n \text{ mod }(2p+1) \neq 0$ (these correspond to Liouville momenta $P = \frac{ib}{4}(2p+1-2n)$). Lian and Zuckerman also describe a general procedure of constructing the corresponding representatives.
\paragraph{Tachyon correlators. }
The simplest cohomology classes that will be of interest to us are ``tachyon'' operators $\mathcal{T}_{1,n}$ of ghost number $1$, corresponding to $n = 1, \dots, p$; they are obtained simply by dressing $\Phi_{1,n}$ with Liouville operators $V_{1-n}$ and ghosts $C\br{C}$ so that we have a primary field of total conformal dimension $0$. Due to identifications in both sectors, similarly constructed representatives for $n = p+1, \dots, 2p$ are idenified with the ones described above up to a constant factor: $\mathcal{T}_{1,n} \sim \mathcal{T}_{1,2p+1-n}$. 

To obtain a non-vanishing correlator, one should saturate the ghost zero modes in the functional integral, i.e. have a specific number of ghost insertions; to this end, when considering correlators of tachyon operators, one may equivalently ``strip away'' the $C$-ghosts and integrate the obtained operator of dimension $1$ over its position. Thus, on general surface, tachyon correlation numbers can be defined as follows:
\begin{equation}
Z^g_{k_1 \dots k_n} := \int \limits_{\br{\mathcal{M}}_{g,n}} Z_{BC} \times \left\langle \prod \limits_{i=1}^n \underbrace{V_{1,-k_i-1} (z_i) \Phi_{1,k_i+1}}_{\Delta = \Delta^L_{1,-k-1} + \Delta^M_{1,k+1} = 1} (z_i)  \right\rangle_{L+MM}
\end{equation}
with $Z_{BC}$ being a ``ghost partition function'' (in fact, a certain correlator that saturates the zero modes) and the integration is over the moduli space of genus $g$ curves with $n$ marked points.

Tachyon correlators are difficult to calculate analytically from worldsheet; integration over the moduli space usually can not be done explicitly. Some reliable analytic results are available only for complex 1-dimensional moduli spaces, i.e. sphere 4-point \cite{Belavin:2005jy} and torus 1-point \cite{Artemev:2022sfi} correlators. Other than that difficulty, for Liouville correlator  calculation involves integrating a product of generic conformal blocks over intermediate Liouville momenta $P$. The integration contour is a real line for small enough external momenta, but is deformed (equivalently, one should account for ``discrete terms'' --- residues from the poles of Liouville structure constants that cross the integration contour \cite{zamzam1996}) if for some $i,j$
\begin{equation} \label{cond}
    |P_i| + |P_j| > \frac{Q}{2}, \text{ or }k_i + k_j < p - \frac{5}{2}.
\end{equation}

\paragraph{Boundary states. } Another type of correlators that will appear in the discussion below are string partition functions with boundaries. In worldsheet approach, they are constructed as follows: first, for a given bordered Riemann surface at each boundary component we impose a conformal boundary condition in all three sectors of worldsheet theory. These conditions can be represented as a ``boundary state''
\begin{equation}
\ket{\sigma; 1,l} = \ket{\sigma}_L \otimes \ket{1,l}_{MM}  \otimes \ket{\text{Gh}} 
\end{equation}
In Liouville, we consider an FZZT state \cite{Fateev:2000ik}, parametrized by a continuous parameter $\sigma$, or boundary cosmological constant $\mu_b \sim \cosh 2\pi b \sigma$; in minimal model, we take the Cardy state \cite{Cardy:1989ir} and $\ket{\text{Gh}}$ in the ghost sector can be described as a ``coherent state'' (see the explicit formula in \cite{polchinski_1998}), solving
\begin{equation}
(c_n + \br{c}_{-n}) \ket{\text{Gh}} = 0,\,(b_n - \br{b}_{-n}) \ket{\text{Gh}} = 0.\,
\end{equation}
We will not need explicit formulas for these states in this work.
The main ``kinematic'' property of Liouville/minimal model physical states is that they satisfy conformal boundary conditions $(L^{L/MM}_n - \br{L}^{L/MM}_{-n}) \left|\sigma /1,l\right\rangle\rangle = 0$. These conditions lead to $\mathcal{Q} \ket{\sigma; k,l}  = -\br{\mathcal{Q}} \ket{\sigma; k,l} $ (see (\ref{brstc})), or $Q_{\text{BRST}} \ket{\sigma; k,l} = 0$ --- boundary states are formally representatives of BRST-cohomology. They, however, include Virasoro descendants at all levels (being constructed from ``Ishibashi states'' \cite{Ishibashi:1988kg}) and thus can not be represented by local operators.

To obtain the string partition function, the CFT correlator described above should be integrated over the moduli space of bordered Riemann surfaces. The result is referred to FZZT-brane amplitude. This calculation is even more technically involved than for the correlators on the closed surfaces; the only examples of explicit worldsheet calculations in minimal string theory the authors are aware of are disk and cylinder partition functions.

In the seminal work of \cite{Seiberg:2003nm}, it is the FZZT-amplitudes, where a state of the form $\ket{\sigma_i; 1,1}$ is associated to each boundary component, that are argued to have the simplest dual ``matrix model'' interpretation (to be discussed in the next section). Moreover, it is argued that more general boundary states of the form $\ket{\sigma; k,l}$  are BRST-equivalent to a certain linear combination of the ones with matter Cardy labels $(1,1)$, so restricting to this case is enough to study all boundary amplitudes.

\subsection{Minimal string matrix model} \label{toprecprelim}
\paragraph{Minimal string and topological recursion. } $(2,2p+1)$ minimal string is usually described using a certain hermitian ``one-matrix model'' after double scaling limit. In the following, we will describe how to obtain perturbative data (correlation numbers and boundary partition functions) in this approach. We will mostly use terminology from a more general framework of topological recursion \cite{Eynard:2007kz}, of which double-scaled matrix models are a particular case. Along the way we will comment on what exactly are the objects that we consider in the ``matrix model'' framework.

The object that we start with is the ``spectral curve'' --- a curve in $\mathbb{C}^2$ defined parametrically by equations
\begin{align}
\begin{cases}\label{MertensSC}
x(z) = z^2 \\
y(z) = \frac{1}{4 \pi} \sin \left(\frac{2}{b^2} \arcsin \pi b^2 z \right)
\end{cases}  \\
 z \in \mathbb{CP}^1,\,b^2 \equiv \frac{2}{2p+1} 
\end{align}
For $b^2$ as above, $y(z)$  can be equivalently written as a Chebyshev polynomial of the first kind $\frac{(-1)^p}{4 \pi}T_{2p+1}(\pi b^2 z)$. 

These equations (more precisely, the 1-form $\omega_{0,1} = y\,dx$) and the bidifferential
\begin{equation}
    B(z_1, z_2) = \frac{dz_1 dz_2}{(z_1 - z_2)^2} \label{bergmank}
\end{equation}
define the recursion kernel
\begin{equation}
    K(z_{n+1}, z) = \frac{\int \limits_{-z}^z B(z_{n+1},\cdot)}{2(y(z) - y(-z)) dx(z)} = \frac{\pi}{(z_{n+1}^2 - z^2)\sin \left(\frac{2}{b^2} \arcsin \pi b^2 z \right)}\frac{d z_{n+1}}{dz}
\end{equation}
and the system of $n$-differentials $\omega_{g,n}(z_1 \dots z_n) \equiv W_{g, n}(z_1, ..., z_n) dz_1 \dots dz_n$. In matrix model, $W_{g,n}$ are related to double-scaled expansion coefficients for correlators of resolvent operators $R_{g,n}(-z_1^2 \dots -z_n^2)$: $R_{g,n} = W_{g,n}/ \prod \limits_{k=1}^n (-2z_k)$. Topological recursion is a relation allowing to calculate $W_{g,n}$ starting from $\omega_{0,1}$ and $\,\omega_{0,2} = B$; it looks as follows
\begin{align} \label{toprecu}
\omega_{g,n+1}(z_1, \dots, z_n,z_{n+1}) =\, &\underset{z=0}{\text{Res}}\;  K(z_{n+1},z) \left(\omega_{g-1,n+2}(z_1, \dots z_n,z,-z) + \right. \nonumber \\
&\left. +\sum \limits_{g_1=0}^g \sum \limits_{J_1 +J_2 = \{z_1 \dots z_n \}} \omega_{g_1, |J|+1} (J_1, z) \omega_{g-g_1, |J_2|+1} (J_2, -z) \right) 
\end{align}
The first few coefficients $W_{g,n}$ are
\begin{equation}
    W_{0, 3} = \frac{1}{z_1^2 z_2^2 z_3^2};\,
    W_{0, 4} = \left( 2\pi^2 - \frac{\pi^2}{2}b^4 + 3 \sum \limits_{i = 1}^{4}\frac{1}{z_i^2}\right) \frac{1}{z_1^2 z_2^2 z_3^2 z_4^2};\,
    W_{1, 1} = \left( \frac{\pi^2}{12} - \frac{\pi^2}{48} b^4 + \frac{1}{8} \frac{1}{z_1^2}\right) \frac{1}{z_1^2}. \nonumber
\end{equation}
In different contexts to conform with the other references it might be convenient to switch to different parametrization, rescaling $z$, or to shift or rescale $x$ and $y$ by a constant; all these transformations do not change the spectral curve in a significant way. For example, one might put equal to $1$ the highest order coefficient for the polynomial $y$: $y = z^{2p+1} + \dots$.

$W_{g,n}(z)$ (or, more precisely, resolvent correlators) are supposed to compute minimal string multi-boundary amplitudes with FZZT boundary conditions and matter labels $(1,1)$ \cite{mertens2021}, where $1 - 2 z^2$ for each boundary has a meaning of boundary cosmological constant \cite{Seiberg:2003nm}. ``Free energies'' $\mathcal{F}_g$ (double-scaled expansion coefficients for logarithm of the partition function) can also be calculated using topological recursion data. For genus $g\geq 2$, they can be defined using the dilaton equation (see (\ref{dilaton}) below), identifying $\mathcal{F}_g \equiv W_{g,0}$. For genus $0$ and $1$, definitions are more involved (see e.g \cite{Eynard:2007kz}).

\paragraph{Correlation numbers.} One can deform the spectral curve and calculate how free energy in different genus changes.  A class of interesting deformations is when $x(z)$ and $y(z)$ are still polynomials of the same parity and the same degree, but their coefficients are varied: $x = z^2 + x_0, y = z^{2p+1} + y_0 z^{2p-1} + \dots$. Then, it is convenient to use the so-called ``KdV coordinates'' $t_k$ in the space of parameters. They are defined as follows (see e.g. \cite{Marshakov:2009mn}, although we use a different normalization):
\begin{equation}
t_{k} = - \frac{1}{2} \frac{g_k}{g_{-2}} \underset{z= \infty}{\text{Res}}\;  \left( x^{1/2+ k-p} y\,dx \right),\,g_k = \frac{(p-k-1)!}{(2p-2k-3)!!} \label{kdvdef}
\end{equation}
To compare with the other approaches, an additional condition $t_{-1} = 0$ can be imposed on the polynomials $x$ and $y$.

There are several ways to calculate free energies in this approach. A convenient method for small genus  makes use of ``Douglas string equation'' (\cite{belzam2009}, \cite{tarn2011}, \cite{beltar2010}); we will not review it here, since we will not use it. Derivatives $\frac{\pd^n \mathcal{F}_g}{\pd t_{k_1} \dots \pd t_{k_n}}$ are referred to as $n$-point ``correlation numbers in KdV frame''.

This is not precisely what we are interested in, however. To compare the results with worldsheet CFT approach, we need to take derivatives with respect to different couplings $\tau_k$, which are related to $t$ via the so-called ``resonance transformations''. Let us parametrize $t_0 = - \frac{1}{2} \frac{g_0}{g_{-2}} u_0^2$. Then, other KdV times are expressed in terms of $\tau_k$ as follows (\cite{tarn2011}):
\begin{equation}
\frac{g_k}{g_{-2}} t_{k} = (2p+1) u_0^{k+2} \sum \limits_{n=1}^{\lfloor \frac{k+2}{2}\rfloor} \sum \limits_{\substack{m_1 \dots m_n = 0 \\ \sum m_l = k+2-2n}} \frac{\tau_{m_1} \dots \tau_{m_n}}{n!} \frac{(2p-2k+2n-5)!!}{(2p-2k-3)!!} \label{taudef}
\end{equation}
That means that $\tau_0 = - 1/2$. As a check of this formula, one can put $\tau_1 \dots \tau_{p-1} = 0$ and $u_0 = 2^{-p/2}$. Then, the couplings $t_{k}^{(0)}$ are such that the spectral curve becomes
\begin{equation}
\begin{cases}
2x = T_2(z) \\
2^{2p} y = T_{2p+1}(z) \\
\end{cases}
\end{equation}
i.e. coincides with the undeformed one (\ref{MertensSC}) up to rescaling and shifts in $x,y$ and $z$.

Parameter $u_0^2$ is proportional to the coupling constant $\mu$ for the exponential interaction term $\mu e^{2b\phi}$ in the worldsheet Liouville CFT. It is known that worldsheet correlation functions have a power-like dependence on $\mu$; this is why we have this simple dependence on $u_0$ above.

Because of resonance transformations,  $n$-th derivatives of free energy with respect to $\tau$ generally can not be expressed only in terms of $n$-point KdV correlation numbers. Lower-point contributions appear when changing $k_i$ if nonlinear terms in resonance transformations become relevant. In the semiclassical limit $b \to 0$, or $p \to \infty$, when $k_i$ is of order $p$, rescaled correlation numbers have a continuous, but non-analytic limit as functions of $\kappa_i \equiv k_i/p$ due to these additional terms. There is, however, analyticity in certain domains in parameter space. Consider the case when (other than $\tau_0$) for all non-zero couplings $\tau_{k_i}, \tau_{k_j}$ $k_i + k_j \geq p-3$. Then it is easy to see that $\forall i,j,l$ $\frac{\pd^2 t_l}{\pd \tau_i \pd \tau_j} = 0$ and the same is valid for higher derivatives. This means that $n$-point correlation number in CFT is just a linear combination of $n$-point correlation numbers in KdV frame and no non-analytic contributions appear. One can note that this condition coincides with the negation of (\ref{cond}), i.e. in worldsheet formulation it means that no discrete terms appear in the Liouville OPE.

In the domain we described, semiclassical limit of correlation numbers is known to coincide with Weil-Petersson volumes, analytically continued to imaginary lengths $l_n = 2\pi i (1 - \frac{2 k_n}{2p+1})$ (\cite{turiaci2021}, \cite{Artemev_2022}). In such a normalization, corrections to the semiclassical answer are polynomials in $b^4$. 

\paragraph{p-deformed volumes. } An observable that is more straightforward to study systematically than correlation numbers described above was introduced in 
\cite{mertens2021} by the name of ``$p$-deformed volumes''. They are given by certain integral transforms of $W_{g,n}(z_1, \dots z_n)$
\begin{equation}\label{VtoW}
\frac{W_{g,n}(z_i)}{\prod \limits_{k=1}^n 2z_k} = \# \prod \limits_{k=1}^n \left( \int \limits_0^\infty d \beta_k\, e^{(1- 2\pi^2 b^4  z_i^2)\beta_k}\int \limits_0^\infty d\lambda_k \,\lambda_k\,\tanh \pi \lambda_k  K_{i\lambda_k}(\beta_k) \right) V_{g,n}(\lambda_1 \dots \lambda_n)
\end{equation}

This integral transform is complicated to perform in general, but one can easily apply it for our purposes, when transforming polynomials in $1/z_i^2$, according to
\begin{equation}
\frac{1}{z^{2(i+1)}} \rightarrow (2 \pi^2 b^4)^{i+1/2}  \frac{\sqrt{\frac{1}{2\pi}}}{2^i \Gamma(i+1) \Gamma(i+3/2)} \prod \limits_{j=1}^i \left(\lambda^2 + \frac{(2j-1)^2}{4}\right) \label{transf}
\end{equation}

\section{Main results}
\subsection{Correlation numbers and p-deformed volumes} \label{volandnum}
An a priori nontrivial observation that can be noted when studying known correlation numbers in low genus is that, in fact, $p$-deformed volumes and correlation numbers in the domain $k_i + k_j>p-3$ are related via analytic continuation: 
\begin{equation}
\left. \frac{\pd^n \mathcal{F}_g}{\pd \tau_{k_1} \dots \pd \tau_{k_n}} \right|_{\tau_0 = - \frac{1}{2}, \tau_1 \dots \tau_{p-1} = 0} \sim V_{g,n} \left( \lambda_j = \frac{i}{2}(2(p-k_j)-1) \right) \label{cornumandpdef}
\end{equation}
Thus, $\lambda$, up to a factor, can be thought of as Liouville momentum for dressing fields: $\lambda = \frac{2P}{b}$. To our knowledge, this relation has not been noted before in the literature. This formula will allow us to systematically examine finite $p$ corrections to tachyon correlators (at least in some domain) in ``matrix model'' using topological recursion. 

To check the validity of the formula (\ref{cornumandpdef}), we performed direct calculation of correlation numbers using ``Douglas string equation'' (see \cite{belzam2009}) and compared them with integral transforms (\ref{VtoW}). Douglas string equation can be used straightforwardly to obtain correlation numbers in genus 0 and 1, which significantly simplify if we only account for linear terms in ``resonance transformations''. In particular, results given in \cite{belzam2009}, \cite{beltar2010}, obtained using this method, can be cross-checked with the formula above; e.g. four-point function on the sphere given in \cite{belzam2009} in the domain $k_i + k_j>p-3$
\begin{align*}
&\left. \frac{\pd^n \mathcal{F}_0}{\pd \tau_{k_1} \dots \pd \tau_{k_4}} \right|_{\tau_0 = - \frac{1}{2}, \tau_1 \dots \tau_{p-1} = 0} = - \frac{1}{2} \left(p(p+1) - \sum \limits_{j=1}^4 (p-k_j)(p-k_j-1) \right) = \\
& =\frac{-1}{2b^4} \left(1 + \frac{3b^4}{4} - b^4\sum \limits_{j=1}^4 (p-k_j-\frac{1}{2})^2\right)
\end{align*}
is reduced to 4-point ``p-deformed volume'' as given in \cite{mertens2021}; in our notations it is
\begin{equation}
V_{0,4}(\vec{\lambda}) = 2\pi^2 + \frac{6\pi^2}{(2p+1)^2} + \frac{8\pi^2}{(2p+1)^2} \sum \limits_{j=1}^4 \lambda_j^2 = 2\pi^2 \left(1 + \frac{3 b^4}{4} + b^4 \sum \limits_{j=1}^4 \lambda_j^2 \right)
\end{equation}
We checked that this relation persists up to 8-point numbers in genus zero and 6-point numbers in genus 1.

Let us describe another way to justify the relation between an integral transform of $n$-differentials ($p$-deformed volumes) and correlation numbers. We would like to find a deformation of the spectral curve such that e.g. $V_{g,1}$ calculate the corresponding variations of free energies and compare this deformation with what happens when we change the couplings $\tau_k$. Deformations of the spectral curve data in terms of small changes of functions $(x(z), y(z)) \in \mathbb{C}((z)) \times \mathbb{C}((z))$ can be parametrized by a $(1, 1)$-form 
\begin{equation}
\Omega = \delta y\, dx - \delta x\, dy
\end{equation}
Here $dx, dy \in \mathbb{C}((z)) dz$ are $(1, 0)$ forms ($1$-forms on the spectral curve) and variations $\delta x$ and $\delta y$ are $(0, 1)$ forms. In terms of coordinates $\{(x_k, y_k)\}_{k}$ such that
\begin{align}
    x(z) = \sum\limits_{k} x_k z^k, \; y(z) = \sum\limits_{k} y_k z^{k},
\end{align}
$\delta x$ and $\delta y$ are given by
\begin{align}
    &\delta x(z) = \sum \limits_{k} \delta x_{k}\, z^k, \; \delta y(z) = \sum \limits_{k} \delta y_{k}\, z^k.
\end{align}

The form $\Omega$ does not depend on whether we vary $x,y$ or both of them and is an invariant characteristic of the deformation; in particular, it is zero if $\delta$ corresponds to reparametrization $z \to f(z)$. According to the general theory \cite{Eynard:2007kz}, if $\Omega$ can be expressed as
\begin{equation}
\Omega (p) = \oint \limits_{(\infty)} \Lambda(q) B(p,q) 
\end{equation}
free energy is deformed as follows
\begin{equation}
\delta \mathcal{F}_g = \oint \limits_{\infty} \omega_{g,1}(z) \Lambda(z)
\end{equation}
Similar formula is valid for $\omega_{g,n}$ themselves and, thus, can be used for higher derivatives: $n$-th derivative of free energy is expressed via $n$-ple contour integral of $\omega_{g,n}$. Equating $\delta \mathcal{F}_g (\Lambda_k) = \sum \limits_{k=0}^{p-1} V_{g,1}(\frac{i}{2} (2p-2k-1))\,\delta \tau_k$, from the explicit transformation formulas (\ref{transf}) one can find  $\Lambda$ as a polynomial (in fact its coefficients ar given by Chebyshev polynomial of the first kind, see (\ref{integerWtoV}))
\begin{equation}
\Lambda(z) = \sum \limits_{k=0}^{p-1} \sum \limits_{i=0}^{p-k} \frac{z^{2i+1}}{\Gamma(i+1)  \Gamma(i+3/2)} \prod \limits_{j=1}^i \left(\frac{(2j-1)^2 - (2p-2k-1)^2}{4} \right) \delta \tau_k
\end{equation}
$\Omega$ then can be calculated, using (\ref{bergmank}). Its coefficients turn out to be proportional to a Chebyshev polynomial of the second kind
\begin{equation}
\Omega(z) = \# \sum\limits_{k=0}^{p-1}  U_{2(p-k-1)}(z)\,dz \,\delta \tau_k
\end{equation}
If we keep $x(z)$ unchanged, to realize this deformation we need to consider non-polynomial deformation of $y$ of the form $\delta y \sim \frac{U_{2(p-k-1)}(z)}{z}$.
The same deformation of spectral curve equation appears in \cite{Seiberg:2003nm}, where it is argued to describe an insertion of tachyon vertex operator $\mathcal{T}_{1,k+1}$ in worldsheet CFT. This is precisely what we wanted to show. In \cite{Seiberg:2003nm}, only one-point function on the disk was considered, so only linear in $\tau_1 \dots \tau_{p-1}$ terms in resonance transformations could be probed. However, they are still nontrivial and in general necessary to account for to obtain this form $\Omega$. In Appendix \ref{app:29}, we rederive this expression for $\Omega$ using explicit formulas for KdV times (\ref{kdvdef}) and resonance transformations (\ref{taudef}), illustrating it on the simplest nontrivial case $p=4$, or $(2,9)$ model. 

If we know the connection between correlation numbers and $p$-deformed volumes, integral transform (\ref{VtoW}) can be motivated by the following idea: LHS, as mentioned before, is supposed to compute the FZZT brane amplitude for minimal string with boundary cosmological constants parametrized by $z_i$, and the first Laplace transform brings us to the partition function on surfaces with boundaries with ``fixed-length'' boundary conditions. These partition functions are decomposed as the correlator of local operators on the punctured surface (which is $V$), to which once-punctured disks are glued (the Bessel function is a one-point function on a disk with these b.c. see e.g. \cite{mertens2021}), summed over a certain complete basis of operators that can be inserted (see some comments on this after (\ref{VtoWsum})). Inverse formula ((\ref{WtoVint}), see below) heuristically describes the opposite operation of gluing punctured disks with bulk operator insertion to surface with boundaries.

\subsection{Comparing spectral curve data for minimal and Virasoro minimal string} \label{varchange}
For the following sections we will use a slightly different parameterization for $(2, 2p+1)$ minimal string spectral curve:
\begin{align}
\begin{cases}
x(z) = T_{2}(\pi b^2 z) = 2 \pi^2 b^4 z^2 - 1\\
y(z) = (-1)^p\, T_{2p+1}(\pi b^2 z) = 
\sin \left(\frac{2}{b^2} \arcsin \left( \pi b^2 z \right) \right)
\end{cases} 
\end{align}
 Bidifferential $B(z_1, z_2)$, needed to define other $n$-differentials, is the same as before:
$
    B(z_1, z_2) = \frac{dz_1 dz_2}{(z_1 - z_2)^2}
$.
Note that in this normalization $W_{g, n}$ differ from the ones defined using $(\ref{MertensSC})$ by a factor $(8\pi^3 b^4)^{2g-2+n}$.
The $p-$deformed volumes are expressed via $W_{g, n}$ using transformation 
\begin{equation}\label{WtoVint}
\frac{ V_{g, n}\left(\lambda_j\right)}{(8\pi^3 b^4)^{2g-2+n}} = \prod \limits_{k=1}^n \left( \frac{\cosh  \pi \lambda_k }{\pi^2 b^2}    \int \limits_0^\infty \frac{d\beta_k}{\beta_k} K_{i\lambda_k}(\beta_k) \int \limits_{-i \infty + \epsilon}^{i \infty + \epsilon}  \frac{dz_k}{2 \pi i}\,  e^{(2 \pi^2 b^4 z_k^2 - 1)\beta_k} \right) W_{g,n}(z_1, \dots , z_n)
\end{equation}
which is inverse to $(\ref{VtoW})$. 

One can integrate over $\beta_k$ (Appendix \ref{app:integral}) and reduce this to the following formula:
\begin{equation}\label{WtoV}
    V_{g, n}(\lambda_j) = (8\pi^3 b^4)^{2g-2+n} \underset{z_1 \dots z_n = 0}{\text{Res}}\;  W_{g, n} (z_1, ..., z_n) \prod \limits_{k=1}^n \frac{dz_k}{2 \pi b^2 \lambda_k} e^{2 \lambda_{k} \arcsin \pi b^2 z_k} 
\end{equation}
For $\lambda_{k} = \frac{i}{2}(2n_k + 1)$ it can be simplified more and we have
\begin{equation}\label{integerWtoV}
   \frac{ V_{g, n}\left(\frac{i}{2}(2 n_{j} + 1)\right)}{(8\pi^3 b^4)^{2g-2+n}} =  \underset{z_1 \dots z_n = 0}{\text{Res}}\;  W_{g, n} (z_1, ..., z_n) \prod \limits_{k=1}^n \frac{ dz_k}{ 2 \pi} \frac{2p+1}{2n_k + 1} (-1)^{n_k} T_{2n_k + 1} \left( \frac{2 \pi z}{2p+1} \right) 
\end{equation}
The form of transformation (\ref{WtoV}) suggests local change of variable:
\begin{equation}\label{stoz}
    z(s) = \frac{1}{\pi b^2 } \sin \pi b s
\end{equation}
1-form $\omega_{0,1}$ and bidifferential $\omega_{0,2}$ in terms of $s$ become
 \begin{align} \label{planchmeasure}
        &\tilde{\omega}_{0, 1}(s)  = 2 \pi b \sin (2 \pi b s) \sin (2 \pi b^{-1} s) ds,\, \nonumber \\
        &B(z(s_1), z(s_2)) = (\pi b)^2 \frac{\cos \pi b s_1 \cos \pi b s_2 }{(\sin \pi b s_1 - \sin \pi b s_2)^2} d s_1 d s_2
\end{align}
Other $\tilde{\omega}_{g, n}(s_1, .., s_n)$ are also given by pullback of $\omega_{g, n}(z_1, ..., z_n)$ under (\ref{stoz}). It follows from the formula (\ref{toprecu}), which defines $\omega_{g,n}$ only in terms of $\omega_{0,1}$ and $B$. To calculate their coefficients $\tilde{W}_{g,n}$ from topological recursion, there is some freedom in choosing an embedding of the spectral curve in $\mathbb{C}^2 \supset (\tilde{x}(s), \tilde{y}(s))$, such that $\tilde{\omega}_{0,1} = \tilde{y}\,d\tilde{x}$. A convenient choice is
\begin{align}
\begin{cases}
    \tilde{x}(s) = s^2\\
    \tilde{y}(s) =y(z(s)) \frac{d x (z(s))}{d \tilde{x}(s)} \\
\end{cases}
\end{align}
The change of variable (\ref{stoz}) allows one to compare topological recursion data for MS and VMS \cite{Collier:2023cyw}. In fact the initial $1$-forms for both theories coincide and (as one can notice from (\ref{WtoV})) $p$-deformed volumes for MS can be obtained from $\tilde{W}_{g,n}$ by simple residue formula
\begin{align}\label{tildeWtoV}
     V_{g, n}(\lambda_j) = (8\pi^3 b^4)^{2g-2+n} \underset{z_k = 0}{\text{Res}}\;  \tilde{W}_{g, n} (s_1, ..., s_n) \prod \limits_{k=1}^n \frac{ds_k}{2 \pi b^2 \lambda_k} e^{2 \pi b \lambda_{k} s_k},
\end{align}
which has exactly the same form as for VMS. The only difference between the recursion data is the nontrivial bidifferential $\tilde{W}_{0, 2}(s)$ for MS in contrast to the standard one for VMS. This form of topological recursion gives us an alternative (to the ``matrix model'') way to calculate perturbative data. We find this relation between two theories very interesting and suggestive. 

Unlike $W(z)$, $\tilde{W}(s)$ --- coefficients for $\tilde{\omega}_{g,n}$  --- do not have an immediate worldsheet interpretation for the minimal string. The same is valid for VMS: its matrix model resolvents are obtained from FZZT amplitudes by a double Laplace transform, first with respect to $s^2$ and then the ``boundary length'' $\beta$. 

One can obtain an intriguing alternative formula for the inversion of ($\ref{tildeWtoV}$); for $\Im s_k < 0$ $\forall k$
\begin{equation}
    \tilde{W}_{g, n}(s_1, ..., s_n) = \frac{1}{(8 \pi^3 b^4)^{2g - 2 - n}}\sum \limits_{n_k = 0}^{\infty} V_{g, n}\left(\frac{i}{2}(2n_k+1)\right) \prod \limits_{k = 1}^{n} (2 \pi b)^2   i \frac{i b}{2} \left( 2 n_k + 1 \right) e^{- 2 \pi s_k \frac{i b}{2}(2 n_k + 1)} \label{VtoWsum}
\end{equation}
Unlike (\ref{VtoW}), this formula represents $\tilde{W}$ as a discrete sum over physical operators (almost all Liouville momenta in the sum correspond to BRST-cohomology classes of \cite{lianzuck}, including not only tachyons, but operators of other ghost numbers)  rather than an integral. One can not exactly take this interpretation at face value: even for tachyons, $V$ coincide with correlation numbers only in a certain region of parameter space. For operators other than tachyons, the relation between $V$ and ``genuine'' correlation numbers is even less clear: it is not known how to introduce deformation parameters so that the corresponding free energy derivatives would compute a correlator with e.g. insertion of several ground ring operators. It would be interesting to better understand this question.
 
\subsection{Intersection theory formula for p-deformed volumes} \label{inttheo}
In this section, we will use the topological recursion described above to find an intersection-theoretic formula for $V_{g,n}$ which is equivalent, but appears to be more suggestive than the one that can be derived from spectral curve (\ref{MertensSC}) and the standard bidifferential. It might be surprising that the same quantities $V_{g,n}$ can be expressed by two different formulas that use tautological classes. However, it can be shown that equivalence of these two formulas is nothing more than a consequence of Virasoro constraints. The technical derivation of this is sketched in Appendix \ref{app: virasoro}\footnote{We are grateful to M. Kazarian for explaining this derivation}.

We apply the technique developed in \cite{Eynard:2011sc} to represent $p$-deformed volumes in terms of intersection of tautological classes on $\mathcal{\overline{M}}_{g, n}$. We recall relevant definitions and properties in Appendix \ref{app:classes}; for a more thorough introduction to intersection theory on $\mathcal{\overline{M}}_{g, n}$, see e.g. \cite{Zvonkine2012AnIT}.

According to \cite{Eynard:2011sc},  $\tilde{W}_{g, n}$ for our topological recursion have the following representation:
\begin{equation}\label{WfromM}
\tilde{W}_{g, n} = 2^{3g - 3 + n} \sum \limits_{j_1, ..., j_n \geq 0} \frac{d \xi_{j_1} }{d s_1} ... \frac{d \xi_{j_n} }{d s_n} \int \limits_{\mathcal{\overline{M}}_{g, n}} \psi_{1}^{j_1}...\psi_{n}^{j_n} e^{\;\sum \limits_{k \geq 0} \tilde{t}_{k} \kappa_{k}  + \sum \limits_{k, l \geq 0} \tilde{B}_{k, l} \delta_{k, l} }
\end{equation}
Integrand in (\ref{WfromM}) is understood as the formal power series in cohomological classes, and integration means pairing of its top degree term with fundamental homology class of $\br{\mathcal{M}}_{g, n}$.  $1$-forms $d \xi_{j}$ are defined by
\begin{equation}
    d \xi_{j} (s) = -\underset{s' = 0}{\text{Res}} \; \tilde{B}(s, s') \frac{(2j - 1)!!}{ 2^{j} s'^{2j + 1}}
\end{equation}
and the times $\tilde{t}_{k}$ are computed from the Laplace transform of the $1$-form $\tilde{\omega}_{0,1} = \tilde{y}\, d \tilde{x}$:
\begin{equation}
    e^{- \sum_{k} \tilde{t}_{k} u^{-k}} = \frac{2 u^{3/2}}{\sqrt{\pi}} \int \limits_{0}^{\infty} e^{-u \tilde{x}(s)} \tilde{\omega}_{0, 1}(s) = 
    4 b \pi^3  e^{\frac{b^2 + b^{-2}}{4} (2 \pi)^2u^{-1} }\frac{(2 \pi)^2u^{-1}/2}{\sinh \left( (2 \pi)^2 u^{-1}/2 \right) }
\end{equation}
Finally, parameters $\tilde{B}_{k, l}$ are defined by the double Laplace transform of the regularized bidifferential (Appendix \ref{app: laplacebergman}):
\begin{align}
    &\frac{e^{- (u_1 + u_2)\sum \limits_{k, l \geq 0} \tilde{B}_{k, l} (u_1 + u_2)^k (u_1 u_2)^l }-1}{u_1 + u_2} =  \frac{1}{2 \pi \sqrt{u_1 u_2}} \int \limits_{s_1 \in \mathbb{R} + i \epsilon} \int \limits_{s_2 \in \mathbb{R}} e^{- \frac{s_1^2}{u_1} -  \frac{s_2^2}{u_2}} \left( \tilde{B}- \overset{\circ}{\tilde{B}} \right) (s_1, s_2)
\end{align}
We see that $\tilde{B}_{k, l} = 0$ for $l \neq 0$ and $\tilde{B}_{k, 0} = (2 \pi)^{2k + 2} \tilde{b}_{k}$, where the first few $\tilde{b}_{k}$'s are given by
\begin{align}
    &\tilde{b}_{0} = \frac{(i b/2)^2}{12}  & \tilde{b}_{3} = -\frac{59(i b/2)^8}{56700}\nonumber \\
    &\tilde{b}_{1} = -\frac{(i b/2)^4}{90} & \tilde{b}_{4} = \frac{232(i b/2)^{10}}{467775}\nonumber \\
    & \tilde{b}_{2} = \frac{8(i b/2)^6}{2835}    & \nonumber
\end{align}
Using (\ref{WtoV}) we find (in terms of $P_{k} = \frac{b}{2}\lambda_k$):
{\allowdisplaybreaks \begin{align}
    &V_{g, n} = \# (8 \pi^3 b^4)^{2g - 2 + n}\sum \limits_{j_1, ..., j_n \geq 0} \prod \limits_{k=1}^{n} \underset{s_k = 0}{\text{Res}}\;  \frac{e^{2 \pi b \lambda_{k} s_k} d\xi_{j_{k}}}{2 \pi b^2 \lambda_{k}} \int \limits_{\mathcal{\overline{M}}_{g, n}} \psi_{1}^{j_1}...\psi_{n}^{j_n} e^{\sum \limits_{k \geq 0} \tilde{t}_{k} \kappa_{k} + \sum \limits_{k \geq 0} \tilde{B}_{k,0} \delta_{k, 0}} = \nonumber \\
    &=\# (8 \pi^3 b^4)^{2g - 2 + n}\sum \limits_{j_1, ..., j_n \geq 0} \prod \limits_{k=1}^{n}  \frac{-1}{b} \frac{(\pi b \lambda_k)^{2 j_k}}{j_k!} \int \limits_{\mathcal{\overline{M}}_{g, n}} \psi_{1}^{j_1}...\psi_{n}^{j_n} e^{\sum \limits_{k \geq 0} \tilde{t}_{k} \kappa_{k} + \sum \limits_{k \geq 0} \tilde{B}_{k,0} \delta_{k, 0}} = \nonumber\\
    &=\# (8 \pi^3 b^4)^{2g - 2 + n}  \int \limits_{\mathcal{\overline{M}}_{g, n}} e^{\frac{c - 13}{24} (2\pi)^2 \kappa_1 -  \sum \limits_{m \geq 1} \frac{B_{2m} (2 \pi)^{4m}}{(2m)(2m)!}\kappa_{2m} + \sum \limits_{k = 1}^{n} (2 \pi)^2 P_k^2 \psi_{k} + \sum \limits_{k \geq 0}\tilde{b}_{k} (2\pi)^{2k+2} \delta_{k, 0}} = \nonumber\\
    &=\# (8 \pi^2 b^2)^{3g - 3 + n}  \int \limits_{\mathcal{\overline{M}}_{g, n}} e^{\frac{c - 13}{24}  \kappa_1 -  \sum \limits_{m \geq 1} \frac{B_{2m}}{(2m)(2m)!}\kappa_{2m} + \sum \limits_{k = 1}^{n}  P_k^2 \psi_{k} + \sum \limits_{k \geq 0}\tilde{b}_{k} \delta_{k, 0}} \label{VfromRR}
\end{align}}
Laplace transform of $d\xi_k$, used in the second line, is calculated in Appendix \ref{app: laplacebergman}.  A few examples of the volumes computed using this formula as well as comments on how to compare them with the other approaches can be found in Appendix \ref{app: volumestable}.

In the formula above, by $c$ we mean the Liouville central charge $c = 13 + 6 b^{2} + 6 b^{-2}$, and $P^2$ parametrize dimensions for external Liouville insertions as $\Delta^L = \frac{c-1}{24} + P^2$. However, the same answers for $p$-deformed volumes are obtained after replacing $b^{2} \to - b^2$,\, $P_k^2 \to - P_k^2$; then the formula is written in terms of parameters of the matter theory. Coefficients $-P_k^2$ before $\psi$-classes  are then related to matter insertions' conformal dimensions in the same way: $\Delta^M = \frac{c_M-1}{24} - P^2$.

\subsection{Discussion on the obtained answer} \label{spec}

 Similarity of this formula with the intersection-theoretic answer for ``quantum volumes'' in VMS
\begin{equation}
V_{g,n}^{\text{VMS}} =  \int \limits_{\mathcal{\overline{M}}_{g, n}} e^{\frac{c - 13}{24}  \kappa_1 -  \sum \limits_{m \geq 1} \frac{B_{2m}}{(2m)(2m)!}\kappa_{2m} + \sum \limits_{k = 1}^{n}  P_k^2 \psi_{k}}
\end{equation}
(the only difference is the additional factor with $\delta_{k,0}$-classes in MS case) suggests a hypothetical relation to worldsheet approach in the spirit of \cite{Collier:2023cyw}. We think that it can be helpful in establishing a more direct understanding (or derivation) of minimal string --- matrix model duality.

What would be a possible explanation for the emergence of these $\delta$-classes? Recall that in VMS, the integrand over moduli space is interpreted as a product of Chern character of ``line bundle of Liouville conformal blocks'' and Todd class of $T\mathcal{M}_{g,n}$. The integrand then is reminiscent of the one in Hirzebruch-Riemann-Roch theorem, which computes dimensions of the space of holomorphic sections of this bundle (more precisely, its Euler characteristic)\footnote{We note in passing that for the minimal string, albeit in the region of parameter space not studied in this paper, correlation numbers on the sphere are also known to be proportional to dimensions of the space of conformal blocks in the \emph{matter} sector \cite{Artemev_2022}.}. 

Line bundles of conformal blocks are well-studied objects in the case of affine Lie algebras \cite{Beauville_1994}. Such bundles are defined over the moduli space of semistable bundles for a given complex curve; over moduli space of curves $\mathcal{M}_{g,n}$ (or its compactification) they then give rise to vector bundles of finite rank via pushforward. Analogous construction for Liouville blocks is less clear: if we think of them as a vector bundle over $\br{\mathcal{M}}_{g,n}$, it would be of infinite rank and would not have a well-defined Chern character. Rigorous construction of line bundles over Teichmuller space $\mathcal{T}_{g,n}$ whose holomorphic sections are Liouville conformal blocks has started to develop recently in \cite{guillarmou2024review}; however, even if this construction can be used to formalize the derivation for VMS, there are still confusing issues with applying HRR theorem to line bundles on non-compact Teichmuller space. We believe that subtleties with treatment of boundary classes and switching between $\mathcal{T}_{g,n}$, $\mathcal{M}_{g,n}$ and $\br{\mathcal{M}}_{g,n}$ in \cite{Collier:2023cyw} need to be understood better. 

For rational CFTs, rigorous definitions of vector bundles of conformal blocks over moduli space of stable curves exist when their rank is finite; in particular, for affine blocks \cite{tsuchiya1989} and in cases like minimal models \cite{Damiolini_2021}. For these examples, on the open part of moduli space the Chern character is given by the same expression as what appears in the VMS formula (up to multiplication by the rank of the bundle). On $\br{\mathcal{M}}_{g,n}$, however, boundary $\delta$-classes are added, with coefficients depending on allowed conformal dimensions in all possible intermediate channels. The general formula for the affine case is given in \cite{marian2016chern2}, but it is complicated; instead of giving that, we will formulate a property from which the coefficients for the boundary classes can be obtained in this context. It characterizes the bundle of conformal blocks $E (\vec{\mu})$ on $\br{\mathcal{M}}_{g,n}$ ($\vec{\mu}$ is a set of irreps of vertex operator algebra associated to marked points) in degeneration limit; namely, it states that pullback  under the attaching map
\begin{equation}
    j: \mathcal{\overline{M}}_{g - 1, n+2} \sqcup \bigsqcup \limits_{h, I} \mathcal{\overline{M}}_{h, I} \times \mathcal{\overline{M}}_{g - h, \bar{I}} \longrightarrow \mathcal{\overline{M}}_{g, n} \nonumber
\end{equation}
decomposes as
\begin{equation}
j^* E (\vec{\mu}) = \mathop{{\bigoplus}} \limits_{\lambda} E (\vec{\mu}, \lambda, \br{\lambda}) \label{bundledecomp}
\end{equation}
i.e. as a direct sum of conformal block bundles on the corresponding moduli spaces, where we associate to two additional marked points all possible irreps $\lambda$ and its dual $\br{\lambda}$. Pullback of the Chern character of conformal block bundle is also given by the sum of characters for bundles with $n+2$ marked points. Although it is best understood for the affine case, this property is very natural to have for a vector bundle of conformal blocks associated to any vertex operator algebra.

Let us try to analyze MS and VMS answers from this perspective. Tautological classes that we work with are of the form $\exp (\dots)$ and start with zero degree cohomology class $1$. Chern characters of vector bundles, on the other hand, start with $\text{rk }E$; thus the best we can assume is that we actually calculate the  (properly regularized\footnote{We should expect this bundle to be of infinite rank. }) ratio of Chern character of this bundle to its rank. In what follows we denote it as ``ch $E$''. Property (\ref{bundledecomp}) for conformal block bundles would lead to  $j^*\text{``ch $E$''}$ decomposing as the sum of such classes for $n+2$ marked points, weighted by ratios of ranks. 

 First, for VMS: if Chern character does not contain boundary classes, as it is defined in \cite{Collier:2023cyw}, the pullback does not lead to anything nontrivial (since $j^*\kappa_k - \kappa_k = j^* \psi - \psi = 0$). If we interpret the integrand in terms of HRR formula for compact moduli space $\br{\mathcal{M}}_{g,n}$, we should have as a second factor a Todd class $\text{td} (T\br{\mathcal{M}}_{g,n})$ with added $\delta$-classes \cite{Eberhardt:2022wlc}:
\begin{equation}
\text{td} (\br{\mathcal{M}}_{g,n}) = \exp \left(-\frac{13}{24}\kappa_1 + \frac{1}{24} \sum \psi_k + \frac{11}{24} \delta_{0,0} - \sum \limits_{m=1}^\infty \frac{B_{2m}}{2m (2m)!} (\kappa_{2m}-\delta_{2m-1,0}) \right)
\end{equation}
Thus, we should add boundary classes to the Chern character as well with the opposite sign, so that the full formula remains the same; what we want to interpret as Chern character of some bundle is
\begin{equation}
\text{``ch $E$''}_{VMS} = \exp \left(\frac{c}{24} \kappa_1 + \sum \limits_{k=1}^n (P_k^2 - \frac{1}{24}) \psi_k - \frac{11}{24} \delta_{0,0} - \sum \limits_{m=1}^\infty \frac{B_{2m}}{2m (2m)!} \delta_{2m-1,0} \right)
\end{equation}
If there is a bundle of conformal blocks (for a certain VOA) associated to these expressions, we expect that when calculating the pullback, we will obtain an expression of the same form, multiplied by a sum over some spectrum $\sum \limits_{\lambda} \exp \left((P^2_\lambda - \frac{1}{24}) (\psi_{n+1} + \psi_{n+2}) \right)$, where $\Delta_\lambda \equiv P^2_\lambda + \frac{c-1}{24}$ is a Virasoro highest weight of VOA irrep $\lambda$ (we used that its dual $\br{\lambda}$ has the same highest weight). Since
\begin{equation}
\delta_{k,0} - j^* \delta_{k,0} = (\psi_{n+1} + \psi_{n+2})^{k+1}
\end{equation}
pullback of $\text{``ch $E$''}$ is given by
\begin{align}
&j^* \text{``ch $E$''}_{VMS} = \exp \left(\frac{c}{24} \kappa_1 + \sum \limits_{k=1}^n (P_k^2 - \frac{1}{24}) \psi_k - \frac{11}{24} \delta_{0,0} - \sum \limits_{m=1}^\infty \frac{B_{2m}}{2m (2m)!} \delta_{2m-1,0} \right)  \times \nonumber \\
&\times \exp \left(- \frac{1}{24} u + \left[\frac{u}{2} + \sum \limits_{m=1}^\infty \frac{B_{2m}}{2m (2m)!} u^{2m} \right]\right) = \exp (\dots) \times \exp \left(-\frac{u}{24}\right) \frac{\exp (u) - 1}{u}
\end{align}
where $u \equiv \psi_{n+1} + \psi_{n+2}$. The last expression can be rewritten in several ways as an integral $\int d\Delta\, \exp ((\Delta - \frac{c}{24}) u)$, similarly to the expected sum over some spectrum.

In the minimal string case there appears an additional factor. It is expressed via the generating function found in Appendix \ref{app: laplacebergman} and looks like
\begin{equation}
\exp (- \sum \limits_{k \geq 0} \tilde{b}_k u^{k+1}) = \frac{b^2}{2} u \sum \limits_{k=0}^\infty \frac{2k+1}{2} \exp \left(- u  \frac{b^2}{4} (k+\frac{1}{2})^2 \right)
\end{equation}
When we multiply by this the answer that was calculated for VMS, additional factor that appears after taking the pullback becomes
\begin{equation}
 \exp \left(-\frac{u}{24} \right) \sum \limits_{k=0}^\infty \frac{2k+1}{2(2p+1)} \left[\exp \left( u  (1-\frac{b^2}{16} (2k+1)^2) \right) - \exp \left(- u \frac{b^2}{16}(2k+1)^2 \right) \right]
\end{equation}
This sum is reminiscent of what we expect --- note that $P_k^2 = - \frac{b^2}{16} (2k+1)^2$ are Liouville momenta for physical operators in minimal string. In accordance with earlier discussion, rational factors can perhaps be interpreted as ratios of ranks. However, we do not know if there is an interpretation for most notable elements of this formula ---  terms with different sign and conformal dimension shifted by $1$.

There are other plausible cohomological interpretations of the expression for MS ``$p$-deformed volumes''; in particular, there is a resemblance to Chiodo classes \cite{Lewanski_2016} that appear when we try to raise the canonical line bundle to rational power. We hope that it will be possible to formulate a more precise explanation for our formula in the future.

\subsection{Recurrence equations: string and dilaton} \label{recequ}
$n$-differentials, calculated using topological recursion, also satisfy simpler linear recurrence relations, commonly referred to as ``string'' and ``dilaton'' equations. They are the consequence of first two Virasoro constraints. Adapted to our case, they are as follows \cite{chaudhuri2019local}
\begin{equation}
\underset{z=0}{\text{Res}}\;  y(z) \frac{\omega_{g.n+1}(z, \vec{z}) }{\prod \limits_{i=1}^n 2 z_i} = - \sum \limits_{k=1}^n dz_k\,\frac{\pd}{\pd (z_k^2)} \left(\frac{\omega_{g.n}(\vec{z})}{dz_k} \frac{1}{\prod \limits_{i=1}^n 2 z_i} \right);\,\vec{z} = (z_1, \dots, z_n)~\text{(string)} \label{string}
\end{equation}
\begin{equation}
 \underset{z=0}{\text{Res}}\;  \Phi(z) \omega_{g.n+1}(z, \vec{z})     = (2g-2+n) \omega_{g.n}(\vec{z}),\,d\Phi = y\,dx ~\text{(dilaton)} \label{dilaton}
\end{equation}
They can be rewritten in a quite simple way for our objects of interest --- $p$-deformed volumes in minimal string theory. Analogous equations were previously derived for usual WP volumes \cite{do2006weilpetersson} and for ``quantum volumes'' in Virasoro minimal string. 

We will derive the string equation for $V_{g,n}$ in detail. 
$W_{g,n}$ is a polynomial in $\frac{1}{z_i^2}$ of the form
\begin{equation}
W_{g,n} = \sum \limits_{i_1 \dots i_n=0} \frac{c_{i_1 \dots i_n}}{z_1^{2(i_1+1)} \dots z_n^{2(i_n+1)}}
\end{equation}
$k$th term in the sum in the RHS of the string equation then reads
\begin{equation}
\mathcal{W}^{(k)}_{g,n} =  \sum \limits_{i_1 \dots i_n=0}  (i_k + 3/2) \frac{c_{i_1 \dots i_n}}{z_1^{2(i_1+1)} \dots z_n^{2(i_n+1)}} \frac{1}{z_k^2}
\end{equation}
After the integral transform  we obtain $p$-deformed volumes
\begin{equation}
V_{g,n} = \sum \limits_{i_1 \dots i_n=0} \tilde{c}_{i_1 \dots i_n} \prod \limits_{k=1}^n \prod \limits_{j_k=1}^{i_k} \left(\lambda_k^2 + \frac{(2j_k-1)^2}{4}\right),\, \tilde{c} := c \cdot \underbrace{\prod \limits_{m=1}^n \frac{(2\pi^2 b^4)^{i_m+1/2} \sqrt{ 
 \frac{1}{2\pi}}}{2^{i_m} \Gamma(i_m+1) \Gamma(i_m+3/2)}}_{\equiv F_{1\dots n}} \label{vseries}
\end{equation}
and on the RHS of the string equation
\begin{equation}
\mathcal{V}^{(k)}_{g,n} = \sum \limits_{i_1 \dots i_n=0} \tilde{c}_{i_1 \dots i_n} \prod \limits_{m=1}^n \prod \limits_{j_m=1}^{i_m} \left(\lambda_m^2 + \frac{(2j_m-1)^2}{4}\right) \cdot \frac{2\pi^2 b^4 \cancel{(i_k+3/2)}}{2(i_k+1) \cancel{(i_k+3/2)}}\left(\lambda_k^2 + \frac{(2i_k+1)^2}{4}\right)
\end{equation}
$\mathcal{V}^{(k)}_{g,n}$ is a polynomial with the following properties:
\begin{itemize}
    \item $\mathcal{V}^{(k)}_{g,n}(\lambda_k = \pm \frac{i}{2}) = 0$, because every summand necessarily has a factor $(\lambda_k^2 + \frac{1}{4})$;
    \item It solves a difference equation 
\begin{equation}
   i\frac{\sin \frac{\pd}{\pd \lambda}}{2\pi^2 b^4} \mathcal{V}_{g,n}^{(k)}(\lambda) \equiv \frac{1}{4\pi^2 b^4} \left(\mathcal{V}^{(k)}_{g,n}(\lambda_k + i) -\mathcal{V}^{(k)}_{g,n}(\lambda_k - i) \right) = i \lambda_k V_{g,n}(\vec{\lambda}) 
\end{equation}
\end{itemize}
The solution of this equation with the initial condition $\mathcal{V}^{(k)}_{g,n}(\lambda_k = \frac{i}{2}) = 0$ in the class of polynomials can be written using the integro-differential operator $\frac{\frac{\pd}{\pd x}}{\sin \frac{\pd}{\pd x}}$ (it is understood as its Maclaurin series and its action on polynomials eventually truncates, so the expression is well-defined)
\begin{equation}
 \mathcal{V}^{(k)}_{g,n}(\tilde{\lambda}) = 2\pi^2 b^4   \int \limits_{i/2}^{\lambda_k} dx \frac{\frac{\pd}{\pd x}}{\sin \frac{\pd}{\pd x}} \left(x V_{g,n}(\vec{\lambda}\, \backslash\, \lambda_k, x)\right)
\end{equation}
Since $V$ is an even function of $\lambda_j$, $\mathcal{V}$ is as well and we also have $\mathcal{V}(-i/2) = 0$. Because of that, as a simple consequence of the difference equation,  for $\lambda_k = \frac{i}{2}(2n+1)$ we can rewrite the answer as a finite sum:
\begin{equation}
\mathcal{V}_{g,n}\left(\lambda_k =  \frac{i}{2}(2n+1)\right) = 4 \pi^2 b^4 \left. \sum \limits_{k=0}^{2n-1-4k > -1} x V_{g,n}(\lambda_k =x) \right|_{x = \frac{i}{2}(2n-1-4k)} 
\end{equation}
In the worldsheet CFT, these $\lambda$ are precisely what corresponds to Liouville momenta of Lian-Zuckerman physical states.

Now, in the LHS we calculate $\text{Res}\; y(z) \omega_{g,n+1}(z,\Vec{z})$, with $y \sim T_{2p+1}(\pi b^2 z)$. From  (\ref{integerWtoV}) we know that calculating residue of $\omega \cdot T_{2n+1}(z)$ is equivalent to transformation (\ref{transf}) with substituting $\lambda = \frac{i}{2}(2n+1)$.
Thus, in the LHS of the string equation after integral transform with respect to $z_1 \dots z_n$ we get $ V_{g,n+1}\left(\frac{i(2p+1)}{2}, \vec{\lambda} \right)$. 

Using parametrization $l = 2\pi b^2 \lambda$ and ``Weil-Petersson'' normalization (when at leading order in $b$ $V_{g,n} = \int \limits_{\br{\mathcal{M}}_{g,n}} \exp (2\pi^2 \kappa_1 + \frac{1}{2} \sum l_i^2 \psi_i)$), string equation assumes the form
\begin{equation}
V_{g,n+1}(2\pi i, \vec{l}) = \sum \limits_{k=1}^n\int \limits_{\pi i b^2}^{l_k} dx\, \frac{2\pi b^2 \frac{\pd}{\pd x}}{\sin 2\pi b^2 \frac{\pd}{\pd x}} \left( x V_{g,n}(x, \vec{l}\, \backslash\, l_k) \right) \label{stringeqf}
\end{equation}
The dilaton equation in the same normalization reads
\begin{equation}
V_{g,n+1}(2\pi i (1-b^2), \vec{l}) - V_{g,n+1}(2\pi i (1+b^2), \vec{l}) = 8 \pi^2 b^2 (2g-2+n) V_{g,n}(\vec{l})
\end{equation}
It is the same as for Virasoro minimal string.

The reason we are interested in these equations is that they are supposed to have a known interpretations in CFT terms, corresponding to the insertion of the simplest kind of operators in the bulk (area operator, for dilaton equation) and on the boundary (length operator, for string equation). Let us make a few comments on that.

Liouville momenta $P_\pm = \frac{l_\pm}{4\pi b} = \frac{i}{2}(b^{-1} \pm b)$, appearing in the LHS of dilaton equation, correspond (respectively) to unit operator $V_{1,1}$ and bulk operator of dimension $1$ $V_{1,-1}$. Insertion of the latter is equivalent to differentiating over Liouville bulk cosmological constant, dependence on which is fixed and power-like (KPZ scaling); thus, for minimal string correlation numbers we have
\begin{equation}
\langle \mathcal{T}_{1,1} \prod \limits_i \mathcal{T}_{1.k_i} \rangle_g \sim \frac{1}{b} \left(\sum \limits_i \frac{b(k_i+1)}{2} - \frac{b^{-1} + b}{2} (2-2g) \right) \langle \prod \limits_i \mathcal{T}_{1.k_i} \rangle_g
\end{equation}
This simpler equation, however, is not valid for $p$-deformed volumes. The correlation number on the LHS is necessarily in the other region of analyticity and one can not analytically continue formulas for $V_{g,n}$ to calculate it. The dilaton equation that is valid for $p$-deformed volumes can be obtained if one imagines the analogous relation with the insertion of matter ``screening operator'' of dimension 1. 
\begin{equation}
\langle \mathcal{T}_{1,-1} \prod \limits_i \mathcal{T}_{1.k_i} \rangle_g \sim \frac{1}{b} \left(\sum \limits_i \frac{b(k_i-1)}{2} - \frac{b^{-1} - b}{2} (2-2g) \right) \langle \prod \limits_i \mathcal{T}_{1.k_i} \rangle_g
\end{equation}
and subtracts it from the equation above. In MS such operator does not ``exist'', being outside of the Kac table, and can only be considered formally. 

The result of substraction reads
\begin{equation}
\langle (\mathcal{T}_{1.1} - \mathcal{T}_{1,-1}) \prod \limits_i \mathcal{T}_{1.k_i} \rangle_g \sim \left(\sum \limits_i 1  - 2 + 2g \right) \langle \prod \limits_i \mathcal{T}_{1.k_i} \rangle_g = \left(2g-2+n \right) \langle \prod \limits_i \mathcal{T}_{1.k_i} \rangle_g 
\end{equation}
which is precisely the dilaton equation. 

As for the string equation, Liouville momentum of additional insertion in the LHS $P = \frac{i b^{-1}}{2}$ corresponds to Liouville operator $V_{1,0} = \exp (b \varphi)$. It is easy to check that one cannot construct a meaningful bulk operator of integer dimension combining it with any minimal model operators from the Kac table. However, it has boundary dimension $1$, as expected of the length operator. Interpretation of the string equation in this terms can also be seen from the form we started with (\ref{string}). In the RHS, we have a sum of derivatives of FZZT boundary amplitude $R = \frac{W}{\prod 2z_i}$ over $z_k^2$. Since $z_k^2$ has a meaning of boundary cosmological constant for each boundary component, differentiating over it is precisely what brings down the length operator.

In ``matrix model''/KdV framework it is known that introduction of the boundary operator can be reduced to a modification of the couplings, corresponding to bulk operators. This modification (as presented e.g. in \cite{Martinec:1991ht}) should be consistent with what we have in the RHS of (\ref{stringeqf}), if we account for (linearized) resonance transformations. It would be interesting to understand how this RHS could appear in the worldsheet approach. This would require to understand formula (\ref{integerWtoV}) in this language and how introducing the length operator interferes with it.

\section{Conclusions} \label{conc}
To conclude, we list some speculations and ideas on possible directions for further study:
\begin{itemize}
    \item Our main aim was to clarify connection between worldsheet and ``matrix model'' formulations for the minimal string like the one of \cite{Collier:2023cyw}; we have not reached that goal yet. Difference between our formula and the one for VMS is an additional factor that includes only $\delta$-classes; we have already speculated on a possible explanation for this modification in the main text. Other than that, it might be possible to interpret this additional factor as a character of some other bundle, tensored with the one of \cite{Collier:2023cyw}. Since $\delta$-classes in a sense ``have support'' on the boundary of moduli space $\pd \br{\mathcal{M}}_{g,n}$, it would be interesting to describe the sections of this bundle as having modified asymptotic behaviour on the boundary. 
    \item The fact that VMS and MS spectral curve data are related in such a simple way needs to be better understood. Perhaps it is connected to the fact that in both cases, the worldsheet CFTs do not have any extra structure other than the Virasoro symmetry (although in the matter sector one deals with either generic or degenerate, respectively, representations of it). The 1-form (\ref{planchmeasure}) has several known interpretations: its coefficient gives the universal Cardy density of states (which is the same as modular kernel for the vacuum character) and the form itself coincides with the Plancherel measure, related to representations of the modular double $\mathcal{U}_{q} (\text{sl}(2,\mathbb{R}))\times \mathcal{U}_{\tilde{q}} (\text{sl}(2,\mathbb{R}))$ \cite{Fan:2021bwt}. It would be interesting to investigate whether this quantum group perspective is useful to understand these two worldsheet theories and their matrix model duals.
    \item Formula (\ref{integerWtoV}) suggests that open string (boundary) states can be expanded in local operators, which allows for an LSZ-type formula connecting open and closed string amplitudes. An analogous idea was already considered for matrix model operators e.g. in \cite{Moore:1991ir}, \cite{Gaiotto:2003yb}. Is there an associated relation in BRST cohomology on the worldsheet between boundary and Lian-Zuckerman states?
    \item An interesting question is whether worldsheet correlators including cohomology classes other than tachyons (e.g. ground ring, when at least the simplest correlators in genus zero can be directly calculated) can be obtained from the ``matrix model'' perspective. KPZ scaling arguments suggest that they might be associated with differentiating over higher KdV couplings; agreement with fusion rules for these correlators, however, would demand them to be incorporated into resonance transformations. It is problematic, since these times have non-positive KPZ dimension.
    
    \item Perhaps a useful way to address a previous point is to consider $(2p+1)$-reduced KP hierarchy instead of $2$-reduced. In topological recursion language, this would amount to swapping functions $x(z)$ and $y(z)$ that define the spectral curve data. This reformulation is interesting, because, apparently, it allows to produce genuine correlation numbers (not just $p$-deformed volumes) from $n$-differentials $\Check{\omega}_{g,n}$ using the same transformation (\ref{integerWtoV}). Thus, in $(2,2p+1)$ MS case resonance transformations are, in a sense, encoded in $x-y$ swap duality relations \cite{Alexandrov:2022ydc}. We checked this statement explicitly on some examples for low $g$ and $n$; a similar result was already found using a related description of MS using Frobenius manifolds \cite{Belavin:2015ffa}. It would also be interesting to understand the cohomological field theory associated to this $xy$-swapped spectral curve.
\end{itemize}

\section{Acknowledgements}
The authors are grateful to M. Kazarian, A. Litvinov, A. Marshakov and B. Feigin for numerous stimulating and insightful discussions. We also would like to thank A. Giacchetto and D. Lewański for useful comments and interest in this work.  The work of A.A. is supported by the Russian Science Foundation grant (project no. 23-12-00333).

\appendix
\section{Spectral curve variations corresponding to CFT couplings: example (2,9)} \label{app:29}
Spectral curve for $(2,9)$ minimal string and general KdV times looks as follows
\begin{equation}
\begin{cases}
x = z^2 + X_0 \\
y = z^9 + Y_0 z^7 + Y_1 z^5 + Y_2 z^3 + Y_3 z
\end{cases}
\end{equation}
KdV times are calculated via (\ref{kdvdef}). Condition $t_{-1} = 0$ yields $Y_0 = \frac{9X_0}{2}$ (this is implied in the following). Other times are expressed as functions of the polynomial's coefficients as follows
\begin{equation}  \cdot
\begin{cases}
\frac{g_{-2} t_0}{g_0} =  
\frac{1}{8} \left(8 Y_1-63 X_0^2\right) \\
\frac{g_{-2} t_1}{g_1} = 
\frac{1}{8} \left(105 X_0^3-20 X_0 Y_1+8 Y_2\right)  \\
\frac{g_{-2} t_2}{g_2} = \frac{1}{128} \left(-945 X_0^4+240 X_0^2 Y_1-192 X_0 Y_2+128 Y_3\right) \\
\frac{g_{-2} t_3}{g_3} = \frac{1}{64} \left(63 X_0^5-20 X_0^3 Y_1+24 X_0^2 Y_2-32 X_0 Y_3\right)  
\end{cases}
\end{equation}
Resonance transformations (\ref{taudef}) for $u_0 = 2^{-p/2} = 1/4$, on the other hand, read
\begin{equation}
 \begin{cases}
\frac{g_{-2} t_0}{9g_0} =  2^{-4}\tau_0 \\
\frac{g_{-2} t_1}{9g_1} = 2^{-6} \tau_1 \\
\frac{g_{-2} t_2}{9g_2} = 2^{-8}(\tau_2 + \frac{3 \tau_0^2}{2}) \\
\frac{g_{-2} t_3}{9g_3} = 2^{-10}(\tau_3 + \tau_0 \tau_1)
\end{cases}   
\end{equation}
Background couplings, when $x$ and $y$ are proportional to Chebyshev polynomials, correspond to $X_0 = \frac{-1}{2}, Y_1 = \frac{27}{16}, Y_2 = \frac{-15}{32}, Y_3 = \frac{9}{256}$. For CFT couplings it means $\tau_0 = -\frac{1}{2},\,\tau_{1\dots 3} = 0$, as expected. Differentiating the equations above, one can now find how variations $\delta\tau_k$ of the couplings near this point are related to variations of polynomial's coefficients $\delta X_0, \delta Y_i$ and calculate the $(1,0)$-form $\Omega$. The answer is
\begin{equation}
\Omega = dz\cdot \frac{9}{512} \left(\delta \tau_0\, U_6 (z) + \delta \tau_1\, U_4(z) + \delta \tau_2\, U_2(z) + \delta \tau_3\, U_0(z) \right)
\end{equation}
with $U$ being Chebyshev polynomials of the second kind, as expected.

\section{Reducing integral transform (\ref{transf}) to residue} \label{app:integral}
We want to simplify the integral transform (\ref{WtoVint}) for $V_{g,n}$;  with respect to each variable $z_i$ we need to calculate the integral of the form
\begin{equation}
I(\lambda) \equiv \frac{\cosh  \pi \lambda}{\pi^2 b^2} \int \limits_0^\infty \frac{dl}{l} K_{i\lambda}(l)  \frac{1}{2 \pi i}\int \limits_{-i \infty + \epsilon}^{i \infty + \epsilon} dz\,  e^{(2 \pi^2 b^4 z^2 - 1)l} W(z),
\end{equation}
where $W(z)$ is a polynomial in $\frac{1}{z^2}$
\begin{equation}
W(z) = \sum \limits_{i = 0} \frac{c_{i}}{z^{2(i+1)}}.
\end{equation}
One has
\begin{align}
     &I(\lambda) = \frac{ \cosh  \pi \lambda}{\pi^2 b^2}  \int \limits_0^\infty \frac{dl}{l} K_{i\lambda}(l)  \frac{1}{2 \pi i}\int \limits_{-i \infty + \epsilon}^{i \infty + \epsilon} dz\,  e^{(2 \pi^2 b^4 z^2 - 1)l} \frac{\partial }{\partial z}\sum \limits_{i = 0} \frac{-c_{i} /(2i+1)}{z^{2i+1}} = \nonumber \\     
     &= 4 b^2 \cosh  \pi \lambda  \int \limits_0^\infty d l\, K_{i\lambda}(l)  \frac{1}{2 \pi i} \int \limits_{-i \infty + \epsilon}^{i \infty + \epsilon} dz\,  e^{(2 \pi^2 b^4 z^2 - 1)l} \sum \limits_{i = 0} \frac{c_{i}/(2i+1) }{z^{2i}} = \nonumber \\
     &= 4 b^2 \cosh  \pi \lambda  \frac{1}{2 \pi i} \int \limits_{-i \infty + \epsilon}^{i \infty + \epsilon} dz \sum \limits_{i = 0} \frac{c_{i}/(2i+1) }{z^{2i}} \int \limits_0^\infty d l\, K_{i\lambda}(l)  e^{-l(1 - 2 \pi^2 b^4 z^2)}
\end{align}
Integral over $l$ is known (see \cite{gradshteyn2007} 6.611, 3); then, 
\begin{align}
     &I(\lambda) = 4 b^2 \cosh  \pi \lambda  \frac{1}{2 \pi i} \int \limits_{-i \infty + \epsilon}^{i \infty + \epsilon} dz \sum \limits_{i = 0} \frac{c_{i}/(2i+1) }{z^{2i}} \frac{\pi i \sin \left( \frac{2  \lambda}{i} \arcsin \pi b^2 z \right) }{\sinh \pi \lambda \sin \left( 2 \arcsin \pi b^2 z \right) } = (\pi b^2 z = \sin \pi b s) \nonumber \\
       &2\pi b \coth  \pi \lambda  \frac{1}{2 \pi i} \int \limits_{-i \infty + \epsilon}^{i \infty + \epsilon} d s  \frac{ i \sin \left( 2 \lambda \pi b \frac{s}{i} \right) }{ \sin \left( \pi b s \right) } \sum \limits_{i = 0} \frac{c_{i}/(2i+1) }{(\frac{1}{\pi b^2} \sin \pi b s)^{2i}} 
\end{align}
Now, we can replace $ i \sin \frac{2\pi \lambda b s}{i} = \sinh (2\pi \lambda b s) = \exp (2\pi \lambda b s) - \cosh (2\pi \lambda b s)$ and rewrite $$\int \limits_{-i \infty + \epsilon}^{i \infty + \epsilon} \cosh (2\pi \lambda b s) \dots = \frac{1}{2} \left(\int \limits_{-i \infty + \epsilon}^{i \infty + \epsilon} - \int \limits_{-i \infty - \epsilon}^{i \infty - \epsilon}\right) \exp (2\pi \lambda b s) \dots,$$ 
since the other factor in the integrand is an odd function. This expression is equal to half-residue of the integrand at $s=0$. Integral of the first term then can also be reduced to summing residues at the poles in the left half-plane; overall, we have
{\allowdisplaybreaks \begin{align}
         &I(\lambda) = \frac{2}{b} \coth  \pi \lambda  \left( \frac{1}{2 \pi i} \int \limits_{-i \infty + \epsilon}^{i \infty + \epsilon} d s \; e^{ 2 \lambda \pi b s}  \sum \limits_{i = 0} \frac{\frac{c_{i}}{(2i+1)} }{(\frac{\sin \pi b s}{\pi b^2})^{2i + 1}} - 
         \frac{1}{2}  \underset{s=0}{\text{Res}}   \; e^{ 2 \lambda \pi b s}  \sum \limits_{i = 0} \frac{\frac{c_{i}}{(2i+1)} }{(\frac{\sin \pi b s}{\pi b^2} )^{2i+1}} ds \right) = \nonumber \\
          &=\frac{\coth  \pi \lambda}{\pi \lambda b^3}   \left( \frac{1}{2 \pi i} \int \limits_{-i \infty + \epsilon}^{i \infty + \epsilon} d s  \cos \left( \pi b s \right) e^{ 2 \lambda \pi b s} W\left(\frac{\sin \pi b s}{\pi b^2}\right) - 
         \frac{1}{2}  \underset{s=0}{\text{Res}}  \cos \left( \pi b s \right) e^{2 \lambda \pi b s}  W\left(\frac{\sin \pi b s}{\pi b^2}\right) ds \right) = \nonumber \\
          &=\frac{\coth  \pi \lambda}{\pi \lambda b^3}   \left( \sum \limits_{n = 0}^{\infty} \underset{s=-\frac{n}{b}}{\text{Res}}  \cos \left( \pi b s \right) e^{ 2 \lambda \pi b s} W\left(\frac{\sin \pi b s}{\pi b^2}\right) ds - 
         \frac{1}{2}  \underset{s=0}{\text{Res}}    \cos \left( \pi b s \right) e^{ 2 \lambda \pi b s} W\left(\frac{\sin \pi b s}{\pi b^2}\right)  ds \right) = \nonumber \\
          &=\frac{1}{\pi \lambda b^3} \coth  \pi \lambda  \left( \sum \limits_{n = 0}^{\infty} (-1)^n e^{-2 \lambda \pi n} - \frac{1}{2} \right) \underset{s=0}{\text{Res}}  \cos \left( \pi b s \right) e^{ 2 \lambda \pi b s} W\left(\frac{\sin \pi b s}{\pi b^2}\right) ds = \nonumber \\        
            &= \frac{1}{2 \pi \lambda b^2} \underset{s=0}{\text{Res}}\;  e^{  2 \lambda \pi b s} W\left(\frac{\sin \pi b s}{\pi b^2}\right) d \frac{\sin \pi b s}{\pi b^2} = \frac{1}{2 \pi b^2 \lambda} \underset{z=0}{\text{Res}}\;  e^{ 2 \lambda \arcsin \pi b^2 z} W(z) d z
\end{align}}

\section{Tautological classes on the moduli space of stable curves}\label{app:classes}
Here we shortly introduce the $\psi$-, $\kappa$- and $\delta$- cohomological classes. One can find more details in a nice exposition \cite{Zvonkine2012AnIT}.

Let $\br{\mathcal{M}}_{g, n}$ be the moduli space of stable $n$-pointed curves of genus $g$. Then there is the universal curve
\begin{align}
    p: \overline{\mathcal{C}}_{g, n} \longrightarrow \overline{\mathcal{M}}_{g, n}
\end{align}
and marked points define the structure sections
\begin{align}
    s_i : \overline{\mathcal{M}}_{g, n} \longrightarrow \overline{\mathcal{C}}_{g, n}.
\end{align}

Let $\Delta \overset{i}{\hookrightarrow} \overline{\mathcal{C}}_{g, n}$ be the set of nodes in the singular fibers. There is a holomorphic line bundle $\mathcal{L}$ over $\overline{\mathcal{C}}_{g, n}$ whose restriction on $\overline{\mathcal{C}}_{g, n} \setminus \Delta$ is cotangent to the universal curve. It defines $n$ line bundles $\mathcal{L}_{i}$ over $\overline{\mathcal{M}}_{g, n}$:
\begin{align}
    \mathcal{L}_{i} = s_i^{*}\mathcal{L}
\end{align}
The $\psi$-classes are their first Chern classes:
\begin{align}
\psi_{i} = c_{1}(\mathcal{L}_{i}) \in H^{2}(\overline{\mathcal{M}}_{g, n})
\end{align}
Let $D_{i}$ be the divisor in $\overline{\mathcal{C}}_{g, n}$ given by $i$th section $s_{i}$. One can twist $\mathcal{L}$ by these divisors and consider the corresponding first Chern class:
\begin{align}
    K = c_1 (\mathcal{L}(\sum \limits_{i = 1}^{n} D_i )) \in H^{2}(\overline{\mathcal{C}}_{g, n})
\end{align}
The $\kappa$-classes are defined to be pushforwards of its powers:
\begin{align}
    \kappa_{m} = p_{*} (K^{m+1}) \in H^{2m}(\overline{\mathcal{M}}_{g, n})
\end{align}
To introduce the $\delta$-classes one needs to consider conormal bundle $N^{\vee}$ on $\Delta$ and define the following classes on $\overline{\mathcal{C}}_{g, n}$:
\begin{align}
    \Delta_{k, l} = i_{*}(c_{1}(N^{\vee})^{k}) \cdot \Delta^{l} \in H^{2k + 4l + 4}(\overline{\mathcal{C}}_{g, n}),
\end{align}
where by $\Delta$ we denote the Poincare dual element in $H^{4}(\overline{\mathcal{C}}_{g, n})$ for the homology class of $\Delta$.
Then the $\delta$-classes are defined to be pushforwards of $\Delta_{k,l}$:
\begin{align}
    \delta_{k, l} = p_{*}(\Delta_{k, l}) \in H^{2k + 4l + 2}(\overline{\mathcal{M}}_{g, n})
\end{align}
In fact $\delta$-classes can be expressed via $\psi$-classes:
\begin{align}
    \delta_{k, l} = \frac{1}{2} j_{*}((\psi_{n+1} + \psi_{n+2})^{k} (\psi_{n+1} \psi_{n+2})^{l}),
\end{align}
where $j$ is the attaching map:
\begin{equation}
    j: \mathcal{\overline{M}}_{g - 1, n+2} \sqcup \bigsqcup \limits_{h, I} \mathcal{\overline{M}}_{h, I} \times \mathcal{\overline{M}}_{g - h, \bar{I}} \longrightarrow \mathcal{\overline{M}}_{g, n}
\end{equation}
In the union $I$ runs over subsets of $\{1, ..., n+2\}$; $\psi_{n+1}$ and $\psi_{n+2}$ denote the $\psi$-classes corresponding to additional 2 punctures.

\section{Integrals and Laplace transforms} \label{app: laplacebergman}
In this appendix, we present some technical calculations needed for the purpose of section \ref{inttheo}. First, Laplace transform of the 1-forms $d\xi_j$ is 
\begin{align}
    &\underset{s = 0}{\text{Res}}\;  \frac{e^{2 \pi b \lambda s} d\xi_{j}}{2 \pi b^2 \lambda} = -   \underset{s = 0}{\text{Res}}\; \underset{s' = 0}{\text{Res}}\; \frac{e^{2 \pi b \lambda s}}{2 \pi b^2 \lambda} \cdot \frac{(2j-1)!!}{2^j s^{' 2j+1}} \tilde{B}(s, s') = \nonumber \\
    &= -\frac{1}{(2 \pi i)^2}\oint \limits_{|s| = 2\epsilon} \oint \limits_{|s'| = \epsilon} \frac{e^{2 \pi b \lambda s}}{2 \pi b^2 \lambda} \cdot \frac{(2j-1)!!}{2^j s^{' 2j+1}} \tilde{B}(s, s') = \nonumber\\ 
    &= -\frac{1}{(2 \pi i)^2}\oint \limits_{|s'| = \epsilon} \oint \limits_{|s| = \epsilon/2} \frac{e^{2 \pi b \lambda s}}{2 \pi b^2 \lambda} \cdot \frac{(2j-1)!!}{2^j s^{' 2j+1}} \tilde{B}(s, s') - \frac{1}{2 \pi i} \oint \limits_{|s'| = \epsilon} \underset{s = s'} {\text{Res}}\; \frac{e^{2 \pi b \lambda s}}{2 \pi b^2 \lambda} \cdot \frac{(2j-1)!!}{2^j s^{' 2j+1}} \tilde{B}(s, s')  = \dots
\end{align}
Here we deformed the integration contour for $s'$-variable and took into account the contribution from the pole at $s' = s$. Since integrand is regular at $s = 0$, the first term vanishes and we proceed 
\begin{align}
    \dots &=- \underset{s' = 0}{\text{Res}}\; \frac{(2j-1)!!}{2^j s^{' 2j - 1}} \underset{z = z(s')}{\text{Res}}\; \frac{e^{2 \lambda \arcsin \pi b^2 z}}{2 \pi b^2 \lambda}\frac{dz dz(s')}{(z - z(s'))^2} =  \nonumber \\
    & = 
    - \underset{s' = 0}{\text{Res}}\; \frac{(2j-1)!!}{2^j s^{' 2j - 1}}dz(s') \left. \frac{\partial}{\partial z}\right|_{ z = z(s')} \frac{e^{2 \lambda \arcsin \pi b^2 z}}{2 \pi b^2 \lambda} = - \underset{s' = 0}{\text{Res}}\; \frac{(2j-1)!!}{2^j s^{' 2j - 1}} \frac{\partial}{\partial s'}\frac{ e^{2 \pi b \lambda s'}}{2 \pi b^2 \lambda} ds' = \nonumber \\
    &= - \frac{(2j-1)!!}{2^j b} \underset{s' = 0}{\text{Res}}\; \frac{e^{2 \pi b \lambda s'}}{s^{' 2j - 1}}  ds' = -\frac{1}{b} \frac{( \pi b \lambda)^{2j}}{j!}
\end{align}
Now, we compute Laplace transform of the bidifferential
{\allowdisplaybreaks
\begin{align}
    &\int \limits_{s_1 \in \mathbb{R} + i \epsilon} \int \limits_{s_2 \in \mathbb{R}} e^{-\tilde{x}(s_1)/u_1 - \tilde{x}(s_2)/u_2}\tilde{B}(s_1, s_2) =
    \int \limits_{\mathbb{R} + i \epsilon} d s_1\, e^{- s_1^2/u_1}\int \limits_{\mathbb{R}} d s_2\, e^{- s_2^2/u_2} (\pi b)^2 \frac{\cos \pi b s_1 \cos \pi b s_2}{(\sin \pi b s_1 - \sin \pi b s_2)^2} = \nonumber\\
    &= \int \limits_{\mathbb{R} + i \epsilon} d s_1\, e^{- s_1^2/u_1} \int \limits_{\mathbb{R}} d s_2\, e^{- s_2^2/u_2} \sum \limits_{k \in \mathbb{Z}} \left( \frac{1}{(s_1 - s_2 + \frac{2k}{b})^2} - \frac{1}{(s_1 + s_2 + \frac{2k - 1}{b})^2} \right)= \nonumber\\
    &=\int \limits_{0}^{\infty}dl\, l \sum \limits_{k \in \mathbb{Z}} \; \int \limits_{\mathbb{R} + i \epsilon} d s_1\, e^{- s_1^2/u_1} \int \limits_{\mathbb{R}} d s_2\, e^{- s_2^2/u_2} \left( e^{il(s_1 - s_2 + \frac{2k}{b})} - e^{il(s_1 + s_2 + \frac{2k - 1}{b})} \right) = \nonumber\\
    & = \sqrt{\pi u_2} \int \limits_{0}^{\infty}dl\, l \sum \limits_{k \in \mathbb{Z}} \; \int \limits_{\mathbb{R} + i \epsilon} d s_1\, e^{- s_1^2/u_1  -\frac{ u_2 l^2}{4} + il (s_1 + \frac{k}{b}) + \pi i k} = \nonumber\\
    &= \pi\sqrt{u_1 u_2} \int \limits_{0}^{\infty} dl\, l e^{-(u_1 + u_2)l^2/4}\sum \limits_{k \in \mathbb{Z}} e^{i \frac{k}{b} (l - \pi b)} = \nonumber \\
    &= 2 \pi^2 b \sqrt{u_1 u_2} \sum \limits_{k \in \mathbb{Z}} \int \limits_{0}^{\infty} dl\, l e^{-(u_1 + u_2)l^2/4}  \delta(l - 2 \pi b (k + 1/2)) = \nonumber\\
    &= 2 \pi \sqrt{u_1 u_2} \cdot \pi b \sum\limits_{k \in \mathbb{Z}} \pi b | k + 1/2 | e^{-(u_1 + u_2)(\pi b (k + 1/2))^2}  \sim  \label{sum}\\ 
    &\sim  \frac{2 \pi \sqrt{u_1 u_2} }{u_1 + u_2} \left(1 + \sum\limits_{m \geq 1}(-1)^{m}\,\frac{(2^{1 - 2m} - 1)  B_{2m}}{m!} ((\pi b)^2 (u_1 + u_2))^{m}  \right)
\end{align}}
The last formula gives an asymptotic expansion for (\ref{sum}). It can be obtained using methods similar to \cite{Mac2018}, or simply by zeta-regularization of divergent series that appear when expanding the summand in series in $(u_1+u_2)$. To calculate coefficients $\Tilde{B}_{k,l}$, one needs to subtract
{\allowdisplaybreaks
\begin{align}
     &\int \limits_{s_1 \in \mathbb{R} + i \epsilon} \int \limits_{s_2 \in \mathbb{R}} e^{- \tilde{x}(s_1)/u_1 - \tilde{x}(s_2)/u_2}\overset{\circ}{\tilde{B}}(s_1, s_2) =
      \int \limits_{\mathbb{R} + i \epsilon} d s_1\, e^{- s_1^2/u_1}\int \limits_{\mathbb{R}} d s_2\, e^{- s_2^2/u_2} \frac{1}{(s_1 - s_2)^2} =\nonumber \\
      &=\pi \sqrt{u_1 u_2} \int \limits_{0}^{\infty} dl\, l e^{-(u_1 + u_2)l^2/4} =  \frac{2 \pi \sqrt{u_1 u_2}}{u_1 + u_2}
\end{align}}

\section{Virasoro constraints} \label{app: virasoro}
Here we show that the possibility of producing different intersection-theoretic expressions by a change of variable on a spectral curve is the consequence of Virasoro constraints. Suppose we have a spectral curve
\begin{align*}
    \begin{cases}
        x = \frac{z^2}{2}\\
        y = \sum\limits_{k = 1}^{\infty} h_{k} \frac{z^{2k-1}}{(2k-1)!!}
    \end{cases},\,  B(z_1 , z_2) = \frac{d z_1 d z_2 }{(z_1 - z_2)^2}
\end{align*}
Let us introduce the expansion coefficients for differentials
\begin{align}
    \omega_{g, n} = \sum\limits_{k_1 ,..., k_n \geq 1} f_{g, k_1, ..., k_n} \prod\limits_{j = 1}^{n}\frac{(2k_j +1)!!}{z_{j}^{2(k_j + 1)}}dz_{j}\\
\end{align}
It is well known that 
\begin{align}\label{fgk}
    &f_{g, k_1, ..., k_n} = \frac{1}{(-h_{1})^{2g - 2 + n}}\int\limits_{\overline{\mathcal{M}}_{g,n}} e^{\sum\limits_{k \geq 1} \tilde{h}_{k} \kappa_{k}}\psi_{1}^{k_{1}}...\psi_{n}^{k_n},\;\;\;\;
    e^{- \sum\limits_{k\geq 1} \tilde{h}_{k} u^{k}} := \sum\limits_{k = 1}^{\infty} \frac{h_k}{h_1} u^{k-1}
\end{align}
Now consider an arbitrary change of coordinates given by an odd function $z = z(s)$. We claim
\begin{align}
    &z^{*}\omega_{g, n} = \sum\limits_{k_1 ,..., k_n \geq 1} \tilde{f}_{g, k_1, ..., k_n} \prod\limits_{j = 1}^{n}d\tilde{\xi}_{k_j}(s_j), \label{changed} \\
    &\tilde{f}_{g, k_1, ..., k_n} =  \frac{1}{(-h_1^{*})^{2g - 2 + n}} \int\limits_{\overline{\mathcal{M}}_{g, n}}  e^{\sum\limits_{k \geq 1} \tilde{h}_{k}^{*} \kappa_{k} + \sum\limits_{k, l \geq 0}\tilde{B}_{k, l}^{*}\delta_{k, l}}\psi_{1}^{k_{1}}...\psi_{n}^{k_n},\\
    &d\tilde{\xi}_{k}(s_j) = \text{Res}_{\tilde{s} = 0}\, B(z(s_j), z(\tilde{s})) \frac{(2k-1)!!}{\tilde{s}^{2k+1}},
\end{align}
where $\tilde{h}^*_k$ and $\tilde{B}^*_{k,l}$ are defined as follows
\begin{align}
&e^{- \sum\limits_{k\geq 1} \tilde{h}_{k}^{*} u^{k}} := \sum\limits_{k = 1}^{\infty} \frac{h_k^{*}}{h_1^{*}} u^{k-1},\;\;\; h^{*}_{k} := \text{Res}_{s = 0}\frac{(2k-1)!!}{s^{2k+1}} y\, dx,
\end{align}
\begin{align}
    &\frac{1 - e^{- (u_1 + u_2 )\sum\limits_{k,l\geq 0}\tilde{B}_{k,l}^{*}(u_{1} + u_{2})^{k}(u_1 u_2 )^{l}}}{u_1 + u_2} := \sum\limits_{k, l \geq 1} B_{k, l}^{*} u_{1}^{k} u_{2}^{l},\,
    B_{k,l}^{*} := \text{Res}_{s = 0} \frac{(2l-1)!!}{s^{2l+1}}d\tilde{\xi}_{k}.
\end{align}
Our claim follows from considering the spectral curve data
\begin{align}
\begin{cases}\label{dsc}
    \tilde{x} = \frac{s^2}{2}\\
    \tilde{y} = \sum\limits_{k = 1}^{\infty} h_{k} \frac{z(s)^{2k-1}}{(2k-1)!!} \frac{z\,dz(s)}{s\, d s} 
\end{cases},\,  \tilde{B}(s_1, s_2) = \frac{d z(s_1) d z(s_2) }{(z(s_1) - z(s_2))^2}
\end{align}
Since $\tilde{y} \,d\tilde{x}$ and $\tilde{B}$ are equal to $z^{*}\omega_{0, 1}$ and $z^* B$, (\ref{dsc}) produces the $n$-differentials in the LHS of (\ref{changed}). Then (\ref{changed}) follows from the  general formula of \cite{Eynard:2011sc}. Now we will rederive this result as a consequence of Virasoro constraints on intersection numbers (\ref{fgk}).

Consider a one-parameter family of changes of coordinates
\begin{align}
    &z = z^{\eta}(s); z^{\eta = 0} = s,\,z^{\eta = 1} = z(s).
\end{align}
It is convenient to use the TR potentials ($\tau_{k}, t_{k}$ are formal variables):
\begin{align}
    &\mathcal{F}^{\eta} = \sum\limits_{g, n \geq 0} \hbar^{2g - 2 + n} \frac{1}{n!} \sum\limits_{k_1, ..., k_n\geq 0} f^{\eta}_{g, k_1, ..., k_n} \tau_{k_1}, ..., \tau_{k_n}\\
    &\mathcal{F} = \sum\limits_{g, n \geq 0} \hbar^{2g - 2 + n} \frac{1}{n!} \sum\limits_{k_1, ..., k_n\geq 0} f_{g, k_1, ..., k_n} t_{k_1}, ..., t_{k_n}.
\end{align}
Introduce a transition matrix $C = C(\eta)$ between bases of odd meromorphic singular 1-forms in a neighborhood of $z = 0$:
\begin{align}
   d \xi^\eta_{k} = \sum\limits_{l = 0}^{k} C_{k, l}(\eta) \frac{(2l+1)!!}{z^{2(l+1)}}dz
\end{align}
Since $\omega_{g,n}^{\eta} = z^{\eta *}\omega_{g,n}$, we have 
\begin{align}\label{fviaffeta}
    &\mathcal{F} = \mathcal{F}^{\eta}(\tau_{k} = \sum\limits_{l = 0}^{k} C_{k, l} t_{l})
\end{align}
Differentiating (\ref{fviaffeta}) with respect to $\eta$ we find
\begin{align} \label{fetaderiv1}
    \dot{\mathcal{F}}^{\eta} = - \sum\limits_{k \geq 0} \sum\limits_{m = 0}^{k}\sum\limits_{l = m}^{k} \dot{C}_{k,l} C^{-1}_{l,m} \tau_{m} \frac{\partial \mathcal{F}^\eta}{\partial \tau_k } 
\end{align}
To prove (\ref{changed}), it is enough to show that
\begin{align}\label{fetans}
    \mathcal{F}^{\eta} = \sum\limits_{g, n \geq 0} \hbar^{2g - 2 + n} \frac{1}{n!} \sum\limits_{k_1 ,..., k_n \geq 1} \frac{\tau_{k_1} \dots \tau_{k_n}}{(-h_1(\eta))^{2g - 2 + n}} \int\limits_{\overline{\mathcal{M}}_{g, n}}  e^{\sum\limits_{k \geq 1} \tilde{h}_{k}(\eta) \kappa_{k} + \sum\limits_{k, l \geq 0}\tilde{B}_{k, l}(\eta)\delta_{k, l}}\psi_{1}^{k_{1}}...\psi_{n}^{k_n}
\end{align}
satisfies the same differential equation. For the derivative of (\ref{fetans}), standard identities on intersection numbers give
\begin{align} \label{fetaderiv2}
    \dot{\mathcal{F}}^{\eta} = \hbar^{-1} \sum\limits_{k \geq 0}\dot{h}_{k}(\eta) \frac{\partial \mathcal{F}^\eta}{\partial \tau_k }  + \frac{1}{2} \sum\limits_{\alpha, \beta \geq 0} \dot{B}_{\alpha, \beta}(\eta) \Big( \frac{\partial^{^2}\mathcal{F}^{\eta}}{\partial \tau_{\alpha} \partial \tau_{\beta}} + \frac{\partial \mathcal{F}^{\eta}}{\partial \tau_{\alpha}} \frac{\partial \mathcal{F}^{\eta}}{\partial \tau_{\beta}} \Big)
\end{align}
Thus we need to prove that RHS of (\ref{fetaderiv1}) and (\ref{fetaderiv2}) are equal. Using (\ref{fviaffeta}) reduces this to equation on $\mathcal{F}$:
\begin{align} \label{feqn}
    \sum\limits_{m}\sum\limits_{l = 0}^{m} \Big( \sum\limits_{k = l}^{m} -C_{k, l} \dot{C}^{-1}_{m, k} \Big) &\Big( t_{l} + \hbar^{-1} \sum\limits_{r = 0}^{l} C^{-1}_{l, r} h_{r}(\eta) \Big) \frac{\partial \mathcal{F}}{\partial t_{m}}\\
    &+ \frac{1}{2} \sum\limits_{k, l \geq 0} \Big( \sum\limits_{\alpha = 0}^{k}\sum\limits_{\beta = 0}^{l} C^{-1}_{k, \alpha} C^{-1}_{l, \beta} \dot{B}_{\alpha, \beta}\Big) \Big( \frac{\partial^{^2}\mathcal{F}}{\partial t_{\alpha} \partial t_{\beta}} + \frac{\partial \mathcal{F}}{\partial t_{\alpha}} \frac{\partial \mathcal{F}}{\partial t_{\beta}} \Big) = 0
\end{align}
Defining
\begin{align}
    v_{m} = - \frac{1}{(2m+1)!!} \sum\limits_{k = 0}^{m}C_{k, 0} \dot{C}^{-1}_{m, k}.
\end{align}
one can check 
\begin{align}
    &- \sum\limits_{k = l}^{m} C_{k, l} \dot{C}_{m, k}^{-1} = \frac{(2m+1)!!}{(2l-1)!!}v_{m - l},\,\sum\limits_{\alpha = 0}^{k} \sum\limits_{\beta = 0}^{l} C^{-1}_{k, \alpha} C^{-1}_{l, \beta} \dot{B}_{\alpha, \beta} = (2k+1)!!(2l+1)!! v_{k + l + 1},
\end{align}
so (\ref{feqn}) can be rewritten as
\begin{align}
    \sum\limits_{m \geq 0} v_{m} \Bigg( \sum\limits_{l \geq 0} & \frac{(2(l + m)+1)!!}{(2l-1)!!} (t_{l} + \hbar^{-1}h_{l}(0)) \frac{\partial \mathcal{F}}{\partial t_{l+m}} + \\
    &+ \frac{1}{2} \sum\limits_{k = 0}^{m - 1} (2k+1)!! (2(m - k)-1)!! \Big( \frac{\partial^{^2}\mathcal{F}}{\partial t_{k} \partial t_{m - k - 1}} + \frac{\partial \mathcal{F}}{\partial t_{k}} \frac{\partial \mathcal{F}}{\partial t_{m - k - 1}} \Big) \Bigg) = 0
\end{align}
This is nothing but a linear combination of Virasoro constraints for Kontsevich-Witten potential with shifted times. Thus, our claim follows from their validity and the procedure above does not lead to novel identities on intersection numbers.

\section{Table of p-deformed volumes} \label{app: volumestable}
Let us present some answers for $p$-deformed volumes obtained with our intersection-theoretic formula. Below they are parametrized by ``geodesic lengths'' $l_k = 4 \pi b P_k $ and are given in Weil-Petersson normalization. We use a condensed notation $m_{\vec{\Lambda}}$ with $\Lambda_1 \geq \Lambda_2 \geq \dots$ to denote a symmetric polynomial in $l_i^2$; e.g. $m_{3,2,1} = \sum \limits_{i\neq j \neq k} l_i^6 l_j^4 l_k^2$.

The simplest way to check the answers for genus zero is to use the explicit formula in \cite{mertens2021} involving the heat capacity for minimal string. Genus 1 answers can be easily reproduced using ``Douglas string equation'' method of \cite{beltar2010}, if we only use linearised resonance transformations.

We also carried out a few checks for $p$-deformed volumes for $g>1$ and zero marked points (equivalently, free energy at genus $g$). They can be alternatively calculated from topological recursion data using the dilaton equation; this was done for the minimal string in \cite{Gregori:2021tvs}. Their formulas for $(g,n) = (2,0), (3,0), (4, 0)$ also agree with our intersection-theoretic answer.
\begin{longtable}{|c|c|}
\hline
      $n$ & $V_{0,n}(\vec{l})$ \\
     & \\[-1em]\hline
     & \\[-1em]
     4 & {$\begin{aligned}2 \pi^2 + \frac{3}{2}\pi^2 b^4 + \frac{1}{2} m_1\end{aligned}$} \\
         & \\[-1em]\hline
         & \\[-1em]
       5 & {$\begin{aligned}10 \pi^4 + 14 \pi^4 b^4 + \frac{13}{2} \pi^4 b^8 + \left(3 \pi^2 + \frac{5}{2} \pi^2 b^4\right) m_1 +
    \frac{1}{2} m_{1,1} + \frac{1}{8} m_2\end{aligned}$}\\   
         & \\[-1em] \hline  
     & \\[-1em]
        6 & {$\begin{aligned} \frac{244}{3} \pi^6 + \frac{493}{3} \pi^6 b^4 + \frac{1651}{12} \pi^6 b^8 + \frac{765}{16} \pi^6 b^{12} +
    \left(26 \pi^4 + 40 \pi^4 b^4 + \frac{229}{12} \pi^4 b^8\right) m_1 +  \\
    + \left(6 \pi^2 + \frac{21}{4} \pi^2 b^4\right) m_{1,1} +
    \left(\frac{3}{2} \pi^2 + \frac{31}{24} \pi^2 b^4\right) m_2 + \frac{3}{4} m_{1,1,1} + \frac{3}{16} m_{2,1} + \frac{1}{48} m_3 \end{aligned}$} \\  
     & \\[-1em]\hline
     & \\[-1em]
        7 & {$\begin{aligned} \frac{2758}{3}\pi ^8 +\frac{7252}{3}\pi ^8 b^4 + \frac{34123}{12}\pi ^8 b^8 + \frac{7109}{4}\pi ^8 b^{12} + \frac{31851}{64}\pi ^8 b^{16}+ \\
    + \left( \frac{910}{3} \pi ^6 +\frac{3995}{6} \pi ^6 b^4 +\frac{4645}{8} \pi ^6 b^8  +\frac{805}{4} \pi ^6 b^{12}  \right) m_1 + \\
    +\left( 80 \pi ^4 + 130 \pi ^4 b^4  + \frac{377}{6} \pi ^4 b^8  \right) m_{1,1} + \left( 20\pi ^4 + \frac{385}{12} \pi ^4 b^4 +\frac{2939}{192} \pi ^4 b^8  \right) m_2 + \\
    +
    \left(15 \pi ^2+\frac{27}{2} \pi ^2 b^4  \right) m_{1,1,1} + 
    \left(\frac{15}{4} \pi ^2 + \frac{10}{3} \pi ^2 b^4 \right) m_{2,1} + \left( \frac{5}{12} \pi ^2  + \frac{35}{96} \pi ^2 b^4  \right) m_3 + \\
    +
    \frac{3}{2}m_{1,1,1,1} + \frac{3}{8} m_{2,1,1} +\frac{3}{32} m_{2,2} + \frac{1}{24} m_{3,1}  + \frac{1}{384} m_4 \end{aligned}$} \\       
     & \\[-1em] \hline
          & \\[-1em] \hline
           & \\[-1em]
    \label{tableg0}
    $n$ & $V_{1,n}(\vec{l})$ \\
    & \\[-1em]\hline
    & \\[-1em]
     & \\[-1em] 
      1 & {$\begin{aligned}\frac{1}{12}\pi^2 + \frac{1}{48}m_1\end{aligned}$} \\
     & \\[-1em] \hline
      & \\[-1em] 
       2 & {$\begin{aligned}\frac{1}{4}\pi ^4 +\frac{5}{24} \pi ^4 b^4 +  \frac{7}{192}\pi ^4 b^8 + \left(\frac{1}{12} \pi ^2  + \frac{1}{24} \pi ^2 b^4 \right) m_1 +\frac{1}{96} m_{1,1} + \frac{1}{192} m_2\end{aligned}$}\\     
    & \\[-1em] \hline
      & \\[-1em]
        3 & {$\begin{aligned}\frac{14}{9}\pi ^6 + \frac{169}{72}\pi ^6 b^4 +\frac{209}{144}\pi ^6 b^8 + \frac{11}{32}\pi ^6 b^{12} + \left( \frac{13}{24} \pi ^4 +\frac{31}{48} \pi ^4 b^4 + \frac{137}{576} \pi ^4 b^8 \right) m_1+\\
    + \left( \frac{1}{8} \pi ^2 + \frac{1}{12} \pi ^2 b^4 \right) m_{1,1}+ \left( \frac{1}{24} \pi^2 + \frac{35}{1152}\pi^2 b^4 \right) m_2\\
    + \frac{1}{96} m_{1,1,1}
    + \frac{1}{192} m_{2,1} 
    + \frac{1}{1152} m_3\\    \end{aligned}$} \\
     & \\[-1em] \hline
      & \\[-1em]
     4 & {$\begin{aligned} \frac{529 \pi ^8}{36} +\frac{1135 \pi ^8 b^4}{36}+\frac{8879 \pi ^8 b^8}{288}+\frac{3125 \pi ^8 b^{12}}{192}+ \frac{3973 \pi ^8 b^{16}}{1024}+\\
    +\left( \frac{187}{36} \pi ^6 + \frac{689}{72} \pi ^6 b^4 +\frac{517}{72} \pi ^6 b^8 +  \frac{421}{192} \pi ^6 b^{12}\right) m_1 + \\
    +\left( \frac{17}{12} \pi ^4 + \frac{47}{24} \pi ^4 b^4 + \frac{317}{384} \pi ^4 b^8   \right) m_{1,1} + \left(\frac{41}{96} \pi ^4 +\frac{175}{288} \pi ^4 b^4 + \frac{613 \pi ^4 b^8 }{2304}\right) m_2 \\
    +\left( \frac{1}{4}\pi^2 + \frac{3}{16} \pi^2 b^4 \right) m_{1,1,1}
    + \left( \frac{1}{12} \pi ^2 + \frac{17}{256} \pi ^2 b^4 \right)m_{2,1} +
    \left( \frac{7}{576} \pi ^2 + \frac{23}{2304} \pi ^2 b^4 \right) m_3\\
    + \frac{1}{64} m_{1,1,1,1}  +  \frac{1}{128} \sum\limits_{j < k} m_{2,1,1} + \frac{1}{384} m_{2,2} + \frac{1}{768} m_{3,1} + \frac{1}{9216} m_4\end{aligned}$} \\      
    \label{tableg1}
     & \\[-1em] \hline
     & \\[-1em] \hline
           & \\[-1em]
    $g$ & $V_{g,0}(\vec{l})$ \\
    & \\[-1em]\hline
    & \\[-1em]
     2 & {$\begin{aligned} \frac{43 \pi ^6}{2160} +\frac{139 \pi ^6 b^4}{8640} +\frac{73 \pi ^6 b^8}{34560}-\frac{17 \pi ^6 b^{12}}{9216} \end{aligned}$} \\
          & \\[-1em] \hline
      & \\[-1em]
     3 & {$\begin{aligned} \frac{176557 \pi ^{12}}{1209600}
    +\frac{2313247 \pi ^{12} b^4}{7257600}
    +\frac{19246753 \pi ^{12} b^8}{58060800}
    +\frac{11651873 \pi ^{12} b^{12}}{58060800} +\\
    +\frac{8154143 \pi ^{12} b^{16}}{132710400}
    -\frac{2202331 \pi ^{12} b^{20}}{619315200}
    -\frac{656431 \pi ^{12} b^{24}}{84934656} \end{aligned}$}\\         & \\[-1em] \hline
      & \\[-1em] 
     4 & {$\begin{aligned}\frac{1959225867017 \pi ^{18}}{493807104000}
    +\frac{1828160015713 \pi ^{18} b^4}{131681894400}
    +\frac{15954711860347 \pi ^{18} b^8}{658409472000} +\\
    +\frac{218467010151779 \pi ^{18} b^{12}}{7900913664000}
    +\frac{94910534310169 \pi ^{18} b^{16}}{4213820620800}
    +\frac{1113449946972163 \pi ^{18} b^{20}}{84276412416000}+\\
    +\frac{519201864600889 \pi ^{18} b^{24}}{101131694899200}
   +\frac{191976434526551 \pi ^{18} b^{28}}{224737099776000}
   -\frac{327865175088253 \pi ^{18} b^{32}}{1198597865472000}-\\
    -\frac{192144632177 \pi ^{18} b^{36}}{1304596316160} \end{aligned}$} \\           
     & \\[-1em] \hline
    \label{tableg}
\end{longtable}
\bibliographystyle{JHEP}
\newpage
\bibliography{mlg2}

\providecommand{\href}[2]{#2}\begingroup\raggedright\begin{thebibliography}{10}

\bibitem{franc1995}
P.~Francesco, P.~Ginsparg and J.~Zinn-Justin, \emph{2d gravity and random matrices}, \href{http://dx.doi.org/10.1016/0370-1573(94)00084-g}{\emph{Physics Reports} {\bf 254} (Mar, 1995) 1–133}.

\bibitem{Moore:1991ir}
G.~W. Moore, N.~Seiberg and M.~Staudacher, \emph{{From loops to states in 2-D quantum gravity}}, \href{http://dx.doi.org/10.1016/0550-3213(91)90548-C}{\emph{Nucl. Phys. B} {\bf 362} (1991) 665--709}.

\bibitem{belzam2009}
A.~A. Belavin and A.~B. Zamolodchikov, \emph{{On Correlation Numbers in 2D Minimal Gravity and Matrix Models}}, \href{http://dx.doi.org/10.1088/1751-8113/42/30/304004}{\emph{J. Phys. A} {\bf 42} (2009) 304004}, [\href{http://arxiv.org/abs/0811.0450}{{\tt 0811.0450}}].

\bibitem{Gregori:2021tvs}
P.~Gregori and R.~Schiappa, \emph{{From Minimal Strings towards Jackiw-Teitelboim Gravity: On their Resurgence, Resonance, and Black Holes}},  \href{http://arxiv.org/abs/2108.11409}{{\tt 2108.11409}}.

\bibitem{Eniceicu:2022dru}
D.~S. Eniceicu, R.~Mahajan, C.~Murdia and A.~Sen, \emph{{Multi-instantons in minimal string theory and in matrix integrals}}, \href{http://dx.doi.org/10.1007/JHEP10(2022)065}{\emph{JHEP} {\bf 10} (2022) 065}, [\href{http://arxiv.org/abs/2206.13531}{{\tt 2206.13531}}].

\bibitem{Witten:1990hr}
E.~Witten, \emph{{Two-dimensional gravity and intersection theory on moduli space}}, \href{http://dx.doi.org/10.4310/SDG.1990.v1.n1.a5}{\emph{Surveys Diff. Geom.} {\bf 1} (1991) 243--310}.

\bibitem{Artemev_2022}
A.~Artemev, \emph{Note on large-p limit of (2,2p+1) minimal liouville gravity and moduli space volumes}, \href{http://dx.doi.org/10.1016/j.nuclphysb.2022.115876}{\emph{Nuclear Physics B} {\bf 981} (aug, 2022) 115876}, [\href{http://arxiv.org/abs/2203.06629}{{\tt 2203.06629}}].

\bibitem{eberhardt20232d}
L.~Eberhardt and G.~J. Turiaci, \emph{{2D dilaton gravity and the Weil-Petersson volumes with conical defects}},  \href{http://arxiv.org/abs/2304.14948}{{\tt 2304.14948}}.

\bibitem{Collier:2023cyw}
S.~Collier, L.~Eberhardt, B.~M\"uhlmann and V.~A. Rodriguez, \emph{{The Virasoro Minimal String}}, \href{http://dx.doi.org/10.21468/SciPostPhys.16.2.057}{\emph{SciPost Phys.} {\bf 16} (2024) 057}, [\href{http://arxiv.org/abs/2309.10846}{{\tt 2309.10846}}].

\bibitem{clessthan12015}
S.~Ribault and R.~Santachiara, \emph{{Liouville theory with a central charge less than one}}, \href{http://dx.doi.org/10.1007/JHEP08(2015)109}{\emph{JHEP} {\bf 08} (2015) 109}, [\href{http://arxiv.org/abs/1503.02067}{{\tt 1503.02067}}].

\bibitem{beltar2010}
A.~Belavin and G.~Tarnopolsky, \emph{{Two dimensional gravity in genus one in Matrix Models, Topological and Liouville approaches}}, \href{http://dx.doi.org/10.1134/S0021364010160137}{\emph{JETP Lett.} {\bf 92} (2010) 257--267}, [\href{http://arxiv.org/abs/1006.2056}{{\tt 1006.2056}}].

\bibitem{Belavin:2005jy}
A.~A. Belavin and A.~B. Zamolodchikov, \emph{{Integrals over moduli spaces, ground ring, and four-point function in minimal Liouville gravity}}, \href{http://dx.doi.org/10.1007/s11232-006-0075-8}{\emph{Theor. Math. Phys.} {\bf 147} (2006) 729--754}.

\bibitem{Artemev:2022sfi}
A.~Artemev and V.~Belavin, \emph{{Torus one-point correlation numbers in minimal Liouville gravity}}, \href{http://dx.doi.org/10.1007/JHEP02(2023)116}{\emph{JHEP} {\bf 02} (2023) 116}, [\href{http://arxiv.org/abs/2210.14568}{{\tt 2210.14568}}].

\bibitem{mertens2021}
T.~G. Mertens and G.~J. Turiaci, \emph{{Liouville quantum gravity -- holography, JT and matrices}}, \href{http://dx.doi.org/10.1007/JHEP01(2021)073}{\emph{JHEP} {\bf 01} (2021) 073}, [\href{http://arxiv.org/abs/2006.07072}{{\tt 2006.07072}}].

\bibitem{Eynard:2011sc}
B.~Eynard, \emph{{Intersection numbers of spectral curves}},  \href{http://arxiv.org/abs/1104.0176}{{\tt 1104.0176}}.

\bibitem{Belavin:1986cy}
A.~A. Belavin and V.~G. Knizhnik, \emph{{Algebraic Geometry and the Geometry of Quantum Strings}}, \href{http://dx.doi.org/10.1016/0370-2693(86)90963-9}{\emph{Phys. Lett. B} {\bf 168} (1986) 201--206}.

\bibitem{Voronov:1988ms}
A.~A. Voronov, A.~A. Roslyi and A.~S. Schwarz, \emph{{Superconformal Geometry and String Theory}}, \href{http://dx.doi.org/10.1007/BF01225506}{\emph{Commun. Math. Phys.} {\bf 120} (1989) 437}.

\bibitem{lianzuck}
B.~H. {Lian} and G.~J. {Zuckerman}, \emph{{Semi-infinite homology and 2D gravity. I}}, \href{http://dx.doi.org/10.1007/BF02099398}{\emph{Communications in Mathematical Physics} {\bf 145} (Apr., 1992) 561--593}.

\bibitem{zamzam1996}
A.~B. Zamolodchikov and A.~B. Zamolodchikov, \emph{{Structure constants and conformal bootstrap in Liouville field theory}}, \href{http://dx.doi.org/10.1016/0550-3213(96)00351-3}{\emph{Nucl. Phys. B} {\bf 477} (1996) 577--605}, [\href{http://arxiv.org/abs/hep-th/9506136}{{\tt hep-th/9506136}}].

\bibitem{Fateev:2000ik}
V.~Fateev, A.~B. Zamolodchikov and A.~B. Zamolodchikov, \emph{{Boundary Liouville field theory. 1. Boundary state and boundary two point function}},  \href{http://arxiv.org/abs/hep-th/0001012}{{\tt hep-th/0001012}}.

\bibitem{Cardy:1989ir}
J.~L. Cardy, \emph{{Boundary Conditions, Fusion Rules and the Verlinde Formula}}, \href{http://dx.doi.org/10.1016/0550-3213(89)90521-X}{\emph{Nucl. Phys. B} {\bf 324} (1989) 581--596}.

\bibitem{polchinski_1998}
J.~Polchinski, \emph{String Theory}, vol.~1 of \emph{Cambridge Monographs on Mathematical Physics}.
\newblock Cambridge University Press, 1998, \href{http://dx.doi.org/10.1017/CBO9780511816079}{10.1017/CBO9780511816079}.

\bibitem{Ishibashi:1988kg}
N.~Ishibashi, \emph{{The Boundary and Crosscap States in Conformal Field Theories}}, \href{http://dx.doi.org/10.1142/S0217732389000320}{\emph{Mod. Phys. Lett. A} {\bf 4} (1989) 251}.

\bibitem{Seiberg:2003nm}
N.~Seiberg and D.~Shih, \emph{{Branes, rings and matrix models in minimal (super)string theory}}, \href{http://dx.doi.org/10.1088/1126-6708/2004/02/021}{\emph{JHEP} {\bf 02} (2004) 021}, [\href{http://arxiv.org/abs/hep-th/0312170}{{\tt hep-th/0312170}}].

\bibitem{Eynard:2007kz}
B.~Eynard and N.~Orantin, \emph{{Invariants of algebraic curves and topological expansion}}, \href{http://dx.doi.org/10.4310/CNTP.2007.v1.n2.a4}{\emph{Commun. Num. Theor. Phys.} {\bf 1} (2007) 347--452}, [\href{http://arxiv.org/abs/math-ph/0702045}{{\tt math-ph/0702045}}].

\bibitem{Marshakov:2009mn}
A.~Marshakov, \emph{{On two-dimensional quantum gravity and quasiclassical integrable hierarchies}}, \href{http://dx.doi.org/10.1088/1751-8113/42/30/304021}{\emph{J. Phys. A} {\bf 42} (2009) 304021}, [\href{http://arxiv.org/abs/0902.4833}{{\tt 0902.4833}}].

\bibitem{tarn2011}
G.~Tarnopolsky, \emph{{Five-point Correlation Numbers in One-Matrix Model}}, \href{http://dx.doi.org/10.1088/1751-8113/44/32/325401}{\emph{J. Phys. A} {\bf 44} (2011) 325401}, [\href{http://arxiv.org/abs/0912.4971}{{\tt 0912.4971}}].

\bibitem{turiaci2021}
G.~J. Turiaci, M.~Usatyuk and W.~W. Weng, \emph{{2D dilaton-gravity, deformations of the minimal string, and matrix models}}, \href{http://dx.doi.org/10.1088/1361-6382/ac25df}{\emph{Class. Quant. Grav.} {\bf 38} (2021) 204001}, [\href{http://arxiv.org/abs/2011.06038}{{\tt 2011.06038}}].

\bibitem{Zvonkine2012AnIT}
D.~Zvonkine, \emph{An introduction to moduli spaces of curves and their intersection theory},  2012.

\bibitem{Beauville_1994}
A.~Beauville and Y.~Laszlo, \emph{Conformal blocks and generalized theta functions}, \href{http://dx.doi.org/10.1007/bf02101707}{\emph{Communications in Mathematical Physics} {\bf 164} (Aug., 1994) 385–419}.

\bibitem{guillarmou2024review}
C.~Guillarmou, A.~Kupiainen and R.~Rhodes, \emph{Review on the probabilistic construction and conformal bootstrap in liouville theory},  2024.

\bibitem{tsuchiya1989}
A.~Tsuchiya, K.~Ueno and Y.~Yamada, \emph{{Conformal Field Theory on Universal Family of Stable Curves with Gauge Symmetries}}, {\emph{Adv. Stud. Pure Math.} {\bf 19} (1989) 459--566}.

\bibitem{Damiolini_2021}
C.~Damiolini, A.~Gibney and N.~Tarasca, \emph{Conformal blocks from vertex algebras and their connections on $\mathcal{M}_{g,n}$}, \href{http://dx.doi.org/10.2140/gt.2021.25.2235}{\emph{Geometry \& Topology} {\bf 25} (Sept., 2021) 2235–2286}.

\bibitem{marian2016chern2}
A.~Marian, D.~Oprea, R.~Pandharipande, A.~Pixton and D.~Zvonkine, \emph{The chern character of the verlinde bundle over the moduli space of stable curves},  2016.

\bibitem{Eberhardt:2022wlc}
L.~Eberhardt, \emph{{Off-shell Partition Functions in 3d Gravity}}, \href{http://dx.doi.org/10.1007/s00220-024-04963-2}{\emph{Commun. Math. Phys.} {\bf 405} (2024) 76}, [\href{http://arxiv.org/abs/2204.09789}{{\tt 2204.09789}}].

\bibitem{Lewanski_2016}
D.~Lewanski, A.~Popolitov, S.~Shadrin and D.~Zvonkine, \emph{Chiodo formulas for the r-th roots and topological recursion}, \href{http://dx.doi.org/10.1007/s11005-016-0928-5}{\emph{Letters in Mathematical Physics} {\bf 107} (Nov., 2016) 901–919}.

\bibitem{chaudhuri2019local}
A.~Chaudhuri, N.~Do and E.~Moskovsky, \emph{Local topological recursion governs the enumeration of lattice points in $\mathcal{M}_{g,n}$},  2019.

\bibitem{do2006weilpetersson}
N.~Do and P.~Norbury, \emph{Weil-petersson volumes and cone surfaces},  2006.

\bibitem{Martinec:1991ht}
E.~J. Martinec, G.~W. Moore and N.~Seiberg, \emph{{Boundary operators in 2-D gravity}}, \href{http://dx.doi.org/10.1016/0370-2693(91)90584-D}{\emph{Phys. Lett. B} {\bf 263} (1991) 190--194}.

\bibitem{Fan:2021bwt}
Y.~Fan and T.~G. Mertens, \emph{{From quantum groups to Liouville and dilaton quantum gravity}}, \href{http://dx.doi.org/10.1007/JHEP05(2022)092}{\emph{JHEP} {\bf 05} (2022) 092}, [\href{http://arxiv.org/abs/2109.07770}{{\tt 2109.07770}}].

\bibitem{Gaiotto:2003yb}
D.~Gaiotto and L.~Rastelli, \emph{{A Paradigm of open / closed duality: Liouville D-branes and the Kontsevich model}}, \href{http://dx.doi.org/10.1088/1126-6708/2005/07/053}{\emph{JHEP} {\bf 07} (2005) 053}, [\href{http://arxiv.org/abs/hep-th/0312196}{{\tt hep-th/0312196}}].

\bibitem{Alexandrov:2022ydc}
A.~Alexandrov, B.~Bychkov, P.~Dunin-Barkowski, M.~Kazarian and S.~Shadrin, \emph{{A universal formula for the $x-y$ swap in topological recursion}},  \href{http://arxiv.org/abs/2212.00320}{{\tt 2212.00320}}.

\bibitem{Belavin:2015ffa}
V.~Belavin and Y.~Rud, \emph{{Matrix model approach to minimal Liouville gravity revisited}}, \href{http://dx.doi.org/10.1088/1751-8113/48/18/18FT01}{\emph{J. Phys. A} {\bf 48} (2015) 18FT01}, [\href{http://arxiv.org/abs/1502.05575}{{\tt 1502.05575}}].

\bibitem{gradshteyn2007}
I.~S. Gradshteyn and I.~M. Ryzhik, \emph{Table of integrals, series, and products}.
\newblock Elsevier/Academic Press, Amsterdam, seventh~ed., 2007.

\bibitem{Mac2018}
R.~McIntosh, \emph{On the asymptotics of some partial theta functions}, \href{http://dx.doi.org/10.1007/s11139-017-9893-6}{\emph{The Ramanujan Journal} {\bf 45} (04, 2018) }.

\end{thebibliography}\endgroup
\end{document}